# Medical Radioisotopes: Clinical Demand, Production Physics, and Untapped Pathways

**Authors:**

Moon J*, Huchthausen C*, Alonso JR, Esfahani SA, Farhadi F, McClain D, Sabet H, Heidari P, Conrad JM, Winklehner D†

**Affiliations:**

- Laboratory for Nuclear Science, Department of Physics, Massachusetts Institute of Technology, 77 Massachusetts Avenue, Building 26, Cambridge, MA 02139, USA (J Moon, C Huchthausen, JR Alonso, D McClain, JM Conrad, D Winklehner)
- Center for Nuclear Medicine and Theranostic Sciences, Department of Radiology, Massachusetts General Hospital, Boston, MA 02114, USA (SA Esfahani, F Farhadi, H Sabet, P Heidari)
- Lawrence Berkeley National Laboratory, 1 Cyclotron Rd, Berkeley, CA, USA (JR Alonso)

**Correspondence:**

D Winklehner (winklehn@mit.edu)

*These authors contributed equally; †Corresponding author

## Summary

Over 40 radioisotopes are currently in use or under development across virtually all fields of modern medicine. Their availability relies on a complex interplay of nuclear physics, radiochemistry, and nuclear medicine. This review provides an isotope-by-isotope map of medically important radioisotopes from the perspective of these disparate yet complementary fields. We address the landscape grouped into three tiers: the established clinical core, including, familiar workhorses such as $^{99m}$Tc, $^{18}$F, $^{177}$Lu; an under-leveraged middle tier including $^{186}$Re, $^{188}$Re, and $^{153}$Sm, and the emerging frontier, radioisotopes with demonstrated but underutilised clinical promise such as $^{225}$Ac, $^{211}$At, $^{67}$Cu, and $^{47}$Sc. We discuss the properties of these radioisotopes with respect to clinical applications and their production pathways, as well as the associated nuclear and radiochemical challenges. We synthesise these disparate perspectives to identify situations in which alleviating production-side bottlenecks may improve access and patient outcomes.

## Introduction

Radioisotopes occupy a vital role in modern medicine where they can serve as imaging agents, therapies, or both. More than 40 radioisotopes are currently used or under development across a wide array of fields: positron emission tomography (PET) and single-photon emission computed tomography (SPECT) imaging, physiological measurement, brachytherapy, external-beam treatment, and targeted radionuclide therapy. Their value arises from the favourable intersection of physical half-life, radiation type and energy, daughter products, chemistry, and biological behaviour. Clinical applications require sufficient specific activity and purity of the radioactive product, properties largely determined by the nuclear reaction used to produce the radioisotope.[1,2] Production meeting these specifications may require a high-flux reactor, a specialised accelerator, enriched target material, or a long-lived generator parent. Irradiation must be followed by processing into clinically usable form and reliable transport to patients far from the production site before decay renders the product unusable.[3]

These three considerations—clinical performance, production, and supply chains—jointly determine whether a radioisotope from a given production route can materially improve patient care. No one property is sufficient. A nuclear reaction yielding large activities may be unusable because it produces a long-lived contaminant that would harm patients. An apparently less clinically favourable radioisotope may succeed because its chemistry, production, and clinical workflow are mature.

In 2024, The Lancet Oncology Series on theranostics mapped radiochemistry trends (clinical and production perspectives), workforce, regulatory access, and LMIC equity (supply chain perspective).[4–7] The present review is the first of two papers in a new Series on medical radioisotopes that builds on the 2024 Series and extends it by surveying medically important radioisotopes isotope by isotope, integrating their nuclear properties, clinical status, and established production methods, and identifies situations where new physics or technology may open realistic alternative pathways that can loosen existing bottlenecks and improve research or patient access. The second paper connects clinical demand and unrealised potential to the supply chain, its vulnerabilities, and the technologies and policies that could widen access worldwide, including in low-income and middle-income countries (McClain D and colleagues, submitted).

### Panel: Search Strategy and Selection Criteria

A per-radioisotope search of Web of Science (all databases), PubMed, and Google Scholar was supplemented by reference tracking from relevant reviews. Eligible records named the radioisotope in the title, abstract, or keywords. Clinical claims required human trial or epidemiological evidence, located using terms including “clinical trial,” “first-in-human,” and phase-designation. Reviews were favoured. Preclinical reports were excluded. Physics and chemistry claims generally required empirical support, with limited exceptions for well-validated simulation studies.

Selection favoured recent and highly cited work, particularly since 2020, without excluding older physics studies whose findings remained current. This review was restricted to English language sources. The production, radiochemical, and clinical evidence relevant to this review is predominantly published in international English language journals, including by groups based in non-English-speaking countries. This restriction was therefore judged unlikely to substantively alter the scientific landscape.

## Conceptual Framework – What Makes a Radioisotope Medically Useful?

A radioisotope becomes medically useful when a series of criteria, spanning nuclear physics to economics, is satisfied. Medically, the isotope’s decay properties and chemistry must permit a safe and clinically meaningful protocol. From the production perspective, the radioisotope must have nuclear generation pathways permitting large-scale creation at sufficient activity and purity. Finally, the radioisotope must be brought to the patient via a supply chain governed by infrastructure, regulation, logistics, demand, and economics. This interplay of the clinical, production, and supply chain perspectives in determining development of a radioisotope’s applications is illustrated for the example of $^{177}$Lu in Figure 1. A typical radioisotope’s journey from production to patient is illustrated in Figure 2.

### Medical perspective

Medically viable radioisotopes are identified by the emissions from the nucleus and their potential for a useful clinical outcome; for example, low-energy photons from $^{125}$I suit uveal melanoma, whereas high-energy alpha particles from $^{223}$Ra suit osteoblastic bone metastases from prostate cancer.

Therapeutic utility also depends on radiochemistry and biochemistry. Can the radioisotope be administered safely and localised to its target? Once on target, does its energy, range, and linear energy transfer (LET) produce the intended damage?

Diagnostic emissions must match an imaging technology. A useful radioisotope might emit photons or positrons suitable for SPECT or PET while minimizing dose that does not contribute to the image.

In either setting, physical and biological half-lives must be long enough for preparation, delivery, administration, and target localization, but short enough to limit residual dose. The entire decay chain determines dose. Short-lived daughters may emit different radiation, while altered chemistry or recoil can change their molecular binding and biodistribution.

Ultimately, medical utility arises from alignment between a radioisotope's nuclear and chemical behaviour and its efficacy in a specific therapeutic or diagnostic situation.[8,9]

## Production perspective

Most medical radioisotopes are produced synthetically, via longer-lived parents in generators or by transmutation using reactors or accelerators. Candidate routes can generally be identified from studied nuclear reaction channels.[2,10]

Target material must be safe, available, and affordable. It must tolerate the thermal and mechanical stress during irradiation and permit rapid post-irradiation processing and delivery within timescales set by the half-life. The reaction must provide adequate yield and purity. A reaction producing abundant $^{225}$Ac, for example, is problematic if it also produces the long-lived impurity $^{227}$Ac.[11]

Production rate is governed by the quantum-mechanical probability of a nuclear reaction—the nuclear cross-section—and the number of interaction attempts, controlled by a beam's intensity on target. A beam of appropriate particle type, energy, and intensity must be attainable with practical reactor or accelerator technology.

## Supply-chain perspective

Successful production does not guarantee patient access. Quality control, transport and packaging, storage, and administration must all occur whilst sufficient activity remains. Especially for short-lived radioisotopes, this means reliable access depends on robust, highly specialised supply chains.

The pathway must be economically sustainable. The many costs of a complex supply chain must be supported by sufficient and predictable clinical demand.

A useful radioisotope reaches patients reliably, affordably, and at a scale supporting routine care.

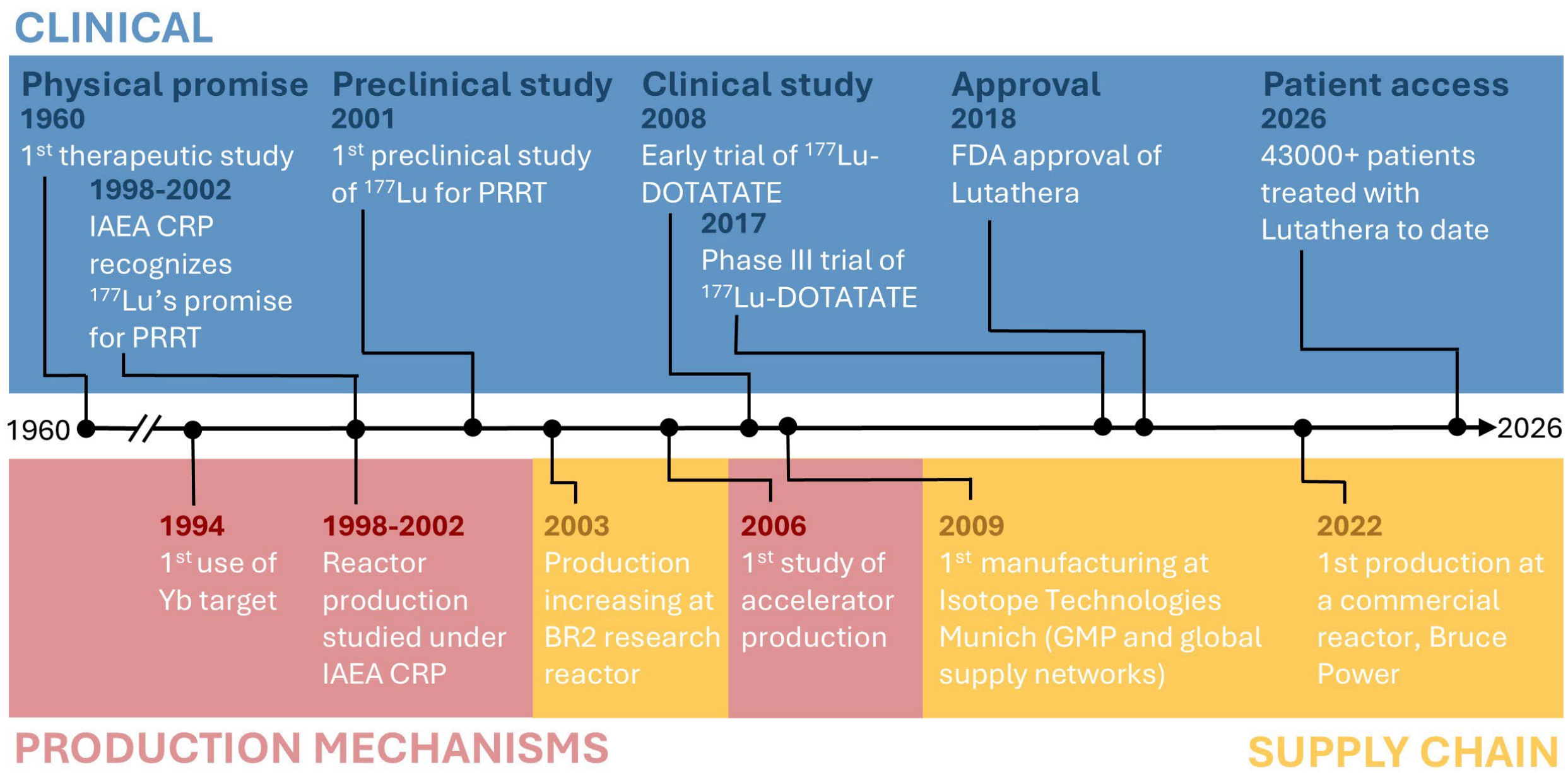

Figure 1. The coevolution of clinical, production, and supply chain aspects that determine patient access to a radioisotope, using $^{177}$Lu-DOTATATE as a case study. CRP=coordinated research project. PRRT=peptide receptor radionuclide therapy. IAEA=International Atomic Energy Agency. GMP=good manufacturing practice.

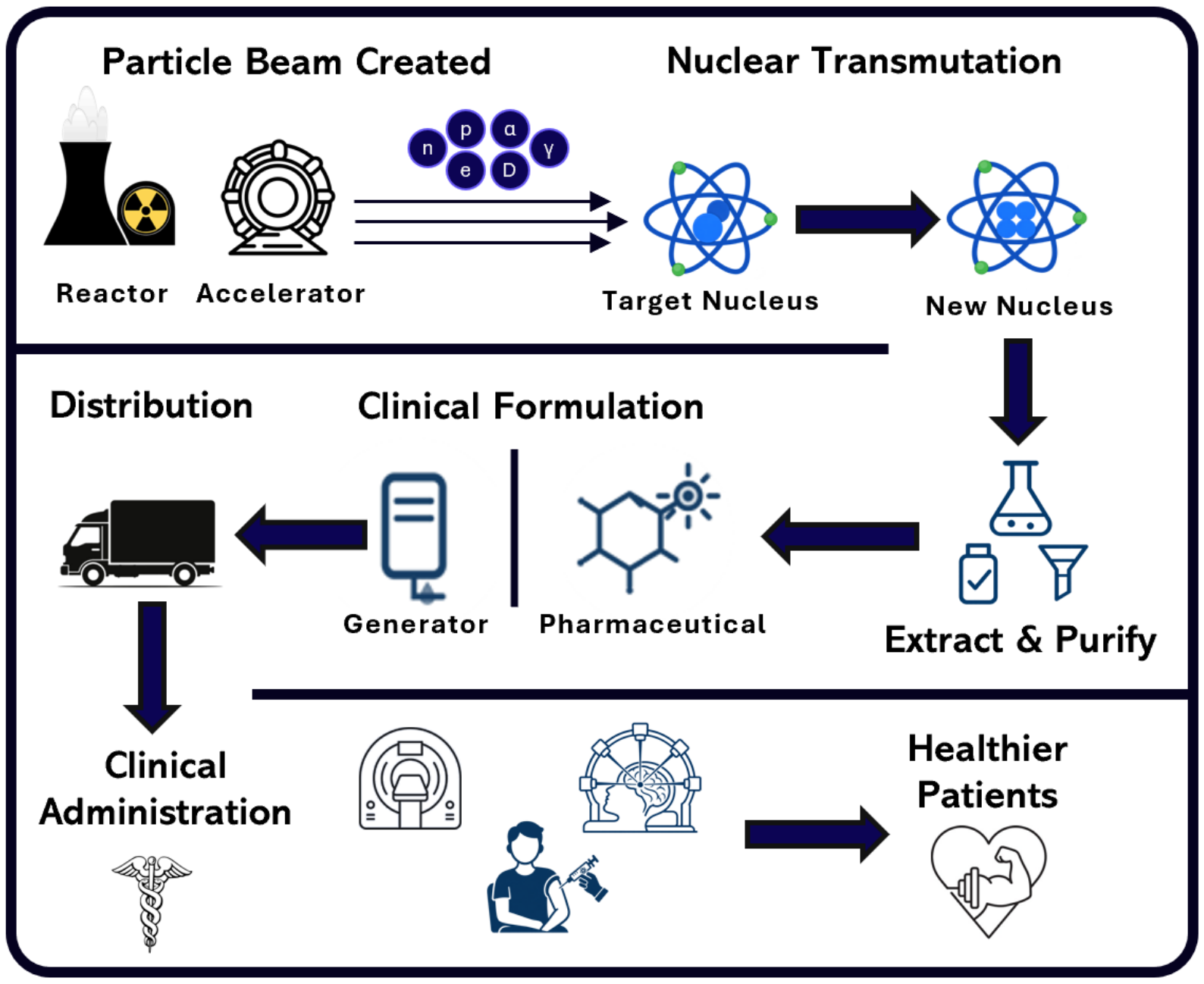

Figure 2. Medical radioisotopes pass through multiple stages on the journey from stable nucleus to patient administration. Smooth interactions between physics, chemistry, distribution logistics, and clinical administration are crucial to optimal patient outcomes.

## The Radioisotope Landscape

The medical radioisotope landscape is not a linear progression from theory to clinical practice. Clinical evidence, production maturity, and supply-chain robustness often advance at different rates. Some radioisotopes matured into well-established radiochemistry and clinical workflows due to the three-way intersection of favourable physical properties, clinical evidence, and scalable production strategies. The under-leveraged middle ground comprises radionuclides with demonstrated, even accepted, clinical value, but a known and available technology has not yet been implemented to broaden access or possible applications. The emerging frontier has attractive physical properties and early clinical evidence but lack scalable production strategies. We explicitly profile several representative radioisotopes below. Figure 3 illustrates all radionuclides considered in several classification schemes. Appendix A presents a full compendium of analysed radioisotopes.

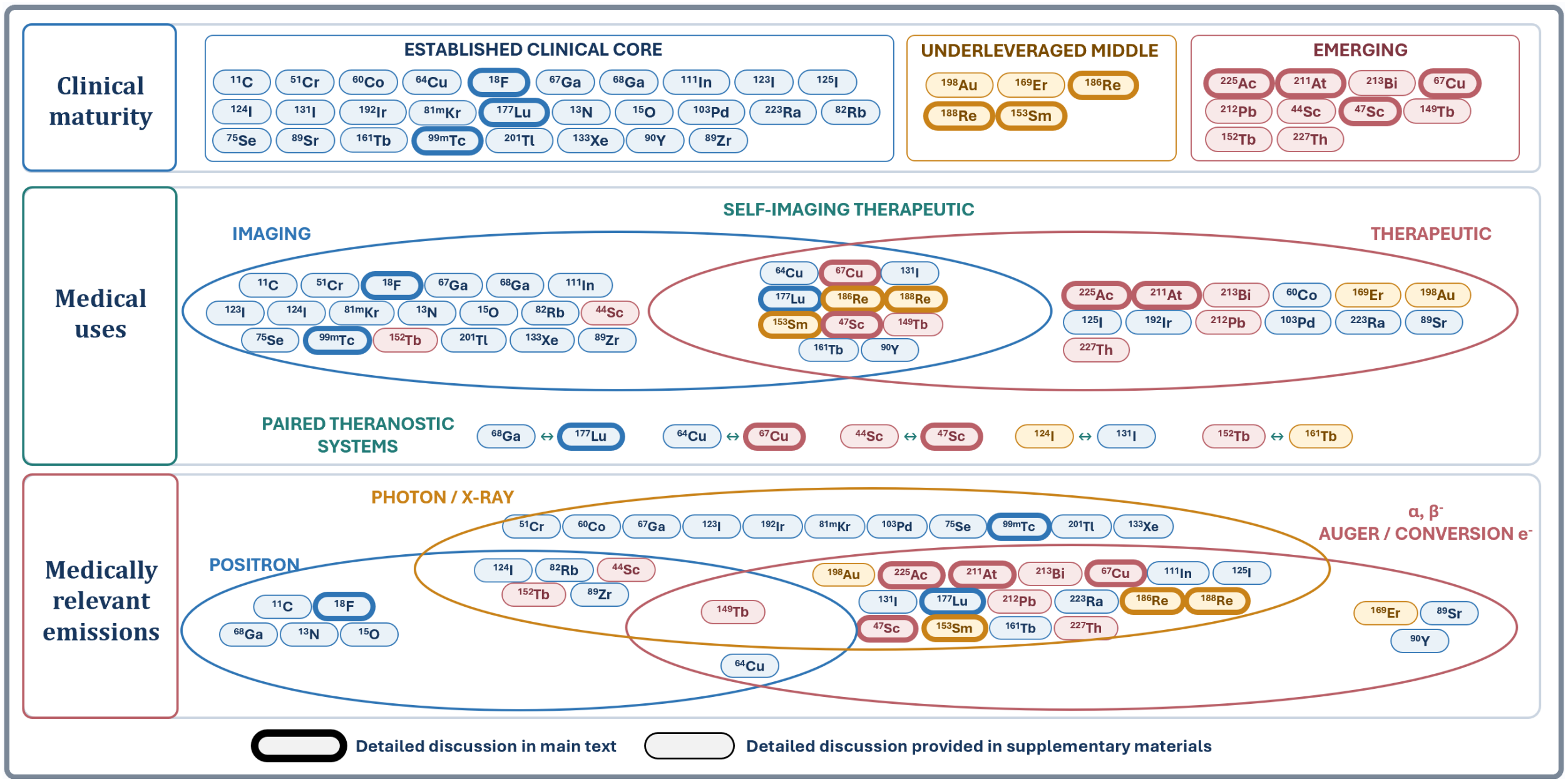

Figure 3. Medical radioisotopes categorised by current clinical maturity, typical use, and medically relevant emissions.

## Clinical Demand

As shown in Table 1, clinical demand differs significantly between the three tiers. The established core has established and growing demand. For example, $^{99m}$Tc supports about 40 million procedures/year and $^{177}$Lu therapies are over 90000/year. The under-leveraged middle is constrained by product form or supply rather than clinical interest. Finally, the emerging frontier shows a trend of increasing therapeutical radiopharmaceutical trials (about 13% per year). However, access is severely limited. For example, $^{225}$Ac-PSMA is available only in 12 countries, while established $^{131}$I is available in about 80. The three tiers thus map to accessibility, decreasing from "Established" to "Emerging". The second paper in this Series discusses how new paradigms in production and distribution can improve accessibility across tiers.

| Tier | Selected Isotopes | Clinical Volume | Reach | Growth Projections | Status |
|---|---|---|---|---|---|
| Established core | $^{99m}$Tc, $^{18}$F, $^{131}$I, $^{177}$Lu | ~40 million $^{99m}$Tc procedures/yr, $^{131}$I ~250 000 therapies/yr, $^{177}$Lu-DOTATATE ~43 000, $^{177}$Lu-PSMA ~48 000 treatments/yr | $^{131}$I in ~80 countries, $^{177}$Lu agents in 32–39 countries | Therapeutic trials increased by approximately 13% per year during 2014–24, $^{177}$Lu-based agents comprised 52.8% of therapeutic trials, and $^{177}$Lu-PSMA indications are moving into earlier treatment lines. | Established and growing |
| Under-leveraged middle | $^{186}$Re, $^{188}$Re, $^{153}$Sm | Established but limited use in bone-pain palliation, radiosynoviorthesis, hepatocellular carcinoma, and non-melanoma skin cancer. | Limited and geographically uneven. | Flat; new indications blocked by specific-activity limits | Constrained by product form or supply-limited |
| Emerging frontier | $^{225}$Ac, $^{211}$At, $^{67}$Cu, $^{47}$Sc | $^{225}$Ac-PSMA ~1000 treatments/yr $^{67}$Cu/$^{47}$Sc trial-scale only | $^{225}$Ac in 12 countries | Therapeutic trials increased by approximately 13% per year during 2014–24, legacy $^{229}$Th-generator $^{225}$Ac production was approximately 63 | Latent and supply-limited |
| | | | | GBq/year, remaining supply inadequate for broad clinical and trial demand | |

Table 1: Clinical demand, volume, reach and projections for the different tiers of radionuclides.

Data sources: $^{99m}$Tc procedure volume/reach from ref [3] ; $^{131}$I, $^{177}$Lu, $^{225}$Ac procedure volumes and $^{225}$Ac reach from refs[7,12] ; $^{67}$Cu/$^{47}$Sc trial-scale only from refs[7,12] ; $^{177}$Lu reach from ref[13]. Therapeutic trials increase, $^{177}$Lu contribution to trials from ref[14], thorium-based $^{225}$Ac production from refs[15,16]

## The Established Clinical Core

Over the past century, a core repertoire of radioisotopes has become embedded in clinical practice. Below we discuss three of these radioisotopes that have become clinical workhorses—$^{18}$F, $^{177}$Lu, and $^{99m}$Tc—as examples demonstrating how physics, chemistry, production, and distribution can come together into a successful system.

### Fluorine-18

$^{18}$F is a positron emitter (96·7%) whose low-energy decay products (mean energy 0·25 MeV)[25] enable high-resolution PET images. Its short 109·8 min half-life allows it to visualise processes with rapid biodistribution kinetics.[25] Fluorine's flexible chemistry lets $^{18}$F combine with a broad variety of tracer molecules,[26] corresponding to a broad variety of processes it can visualise. An $^{18}$F-tagged glucose analogue, FDG, is the standard for imaging numerous cancers and metabolic disorders.[26,27] Additional approved uses include imaging bone turnover, amyloid or tau aggregation, myocardial perfusion, and many more.[28,29]

A convenient production route has made it possible for $^{18}$F to become invaluable in the clinic. Compact medical cyclotrons, small enough to be on-site at a hospital, produce $^{18}$F by irradiating highly enriched water ($H_2{}^{18}O$) with protons. On-site production overcomes the logistical challenge of $^{18}$F's clinically valuable short half-life. The expensive enriched water can be reused after aqueous $^{18}$F is separated,[26] making production sufficiently economical. Automated nucleophilic labelling modules convert the fluoride into FDG or other compounds, useable immediately for imaging.

### Lutetium-177

$^{177}$Lu has recently become a workhorse therapeutic radioisotope in peptide receptor radionuclide therapy: $^{177}$Lu-based radiopharmaceuticals are US Food and Drug Administration (FDA) approved to treat prostate cancer (Pluvicto) and neuroendocrine tumours (Lutathera). It decays with a half-life of 6·65 d by emitting a moderate-energy $\beta^-$ particle (mean 0·134 MeV, maximum 0·498 MeV) that travels far enough in tissue to deposit ionising radiation across a diameter of about 20 cells.[25] Its γ-rays (113 and 208-keV)[25] permit post-treatment SPECT imaging and dosimetry,[30] and due to chemical similarity $^{177}$Lu is often paired with the established PET radioisotope $^{68}$Ga.[31]

Over the last two decades (Figure 1) $^{177}$Lu production has been developed and scaled to enable the roll-out of $^{177}$Lu-based radiopharmaceuticals. In contrast with local $^{18}$F production, $^{177}$Lu production is heavily centralised at reactors. Irradiating enriched $^{176}$Lu in reactors produces $^{177}$Lu in large quantities, but the carrier-added product has specific activity too low for many treatments.[24] Irradiating $^{176}$Yb in reactors to obtain high specific activity $^{177}$Lu has thus become standard.[24] Although the reaction occurs at a very low rate (σ=2·85 b), medium- and high-flux reactors make this route sufficiently economical for widespread clinical use.

### Technetium-99m

$^{99m}$Tc use has risen to approximately 40 million procedures annually, representing about 80% of nuclear-medicine examinations. This is because its decay characteristics are uniquely matched to clinical needs. $^{99m}$Tc emits a 140·5 keV photon during its decay,[25] an energy well-suited to planar and SPECT imaging. Its 6 h half-life permits radiopharmaceutical preparation,[25] uptake, and imaging while limiting patient radiation exposure. Major applications include bone, myocardial-perfusion, renal, hepatobiliary, and lung imaging.[10]

The widespread use of $^{99m}$Tc has been enabled by $^{99}$Mo/$^{99m}$Tc generators. Although production of $^{99m}$Tc's parent radioisotope $^{99}$Mo is centralised, conventionally obtained by thermal-neutron fission of $^{235}$U in reactors,[10] the half-life of $^{99}$Mo is long enough (65·9 h)[25] for it to be shipped, enabling distribution of the short-lived daughter. $^{99m}$Tc is chemically separated from this generator on-site for hospital use.

These examples show that radioisotopes with properties matched to a clinical problem can become established only insofar as their production becomes established. However, even mature production can have vulnerable supply chains. $^{18}$F requires access to a medical cyclotron. $^{177}$Lu and $^{99m}$Tc rely on adequate and predictable reactor capacity. New technologies may thus be valuable even for established radioisotopes if they improve capacity or increase supply robustness through regional production, if they prove their benefit to the supply chain without sacrificing the purity and cost of the existing production route.

## The Under-Leveraged Middle

A third category comprises radioisotopes with demonstrated clinical value and established production that are nonetheless constrained by their current production strategies. For example, some clinical uses for $^{186}$Re are well established, but the expansion to other clinical uses is restricted by availability of high-specific-activity forms that require a different production strategy. The example of $^{188}$Re demonstrates the fragility created by dependence on a small number of very-high-flux reactors. The use of $^{153}$Sm is limited both by lack of high specific activity and distribution challenges. The underleveraged opportunities for such radioisotopes are technologies that do not simply increase production volume, but enable a different product or a different supply architecture.

### Rhenium-186

$^{186}$Re decays predominantly by emitting $\beta^-$ particles (mean 0·35 MeV, maximum 1·08 MeV)[25] with a maximum range in soft tissue of 4·5 mm.[17] $^{186}$Re sulphide is established for radiosynoviorthesis—irradiating overactive cells that cause joint inflammation—of medium-sized joints, which match its $\beta^-$ range.[16] It has also demonstrated palliation of bone pain from metastases.[27] Its 137 keV photon (9·5% of decays)[25] allows concurrent SPECT imaging.[17]

$^{186}$Re's range makes it promising to treat tumours of mm to cm scales. Rhenium is chemically similar to technetium, and $^{186}$Re radiopharmaceuticals could be designed analogous to well-developed $^{99m}$Tc tracers that selectively accumulate in tumours.[17] $^{186}$Re's 3·72 d half-life alternatively makes it suitable for targeting by molecules like antibodies with slow biodistribution.[17] However, its established production route is neutron capture on $^{185}$Re, yielding a carrier-added product. This is suitable for bone and joint applications but limits its use in targeting structures like tumours with a limited number of receptors or antigens, which requires high specific activity.

### Rhenium-188

$^{188}$Re decays by emitting $\beta^-$ particles (mean 0·78 MeV, maximum 2·12 MeV) with a half-life of 17 h.[25] $^{188}$Re–HEDP has demonstrated use for palliation of metastatic bone pain, and transarterial $^{188}$Re–lipiodol has shown efficacy against hepatocellular carcinoma.[28,29] Recent work on topical formulations has shown promise for basal and squamous-cell carcinoma.[30] $^{188}$Re's 155 keV photon (15% of decays)[25] permits concurrent SPECT imaging and dosimetry.

Like $^{186}$Re, $^{188}$Re is a candidate for development of therapeutic radiopharmaceuticals, but it suffers a different bottleneck, because its established production using $^{188}$W/$^{188}$Re generators already yields a high-specific-activity product. The $^{188}$W parent is produced in reactors by successive neutron capture on enriched $^{186}$W. The second capture competes with intermediate $^{187}$W decay, so exceptionally high neutron flux is required, restricting production to only a few reactors worldwide.[31,32]

### Samarium-153

$^{153}$Sm emits $\beta^-$ particles (mean 0·233 MeV, maximum 0·803 MeV)[25] whose dose is well localised with a range in tissue of 0·5-3 mm. Its 103 keV photon (29% of decays)[25] permits post-treatment SPECT imaging.[33] Its current primary use is for palliative care in bone metastases. When administered as $^{153}$Sm-EDTMP it localises exceptionally well to calcium-phosphate compound sites in bone and relieves painful lesions.

High specific activity $^{153}$Sm shows promise for broader applications, having demonstrated receptor-specific tumour uptake, SPECT visualisation, and cytotoxicity in preclinical studies.[34] However, like $^{186}$Re, its routine production, irradiating $^{152}$Sm in reactors, allows high yields but a low-specific-activity product. $^{153}$Sm's short 1·93 d half-life is a significant logistical challenge for distribution far from centralised reactor locations.[25]

Developing accelerator technologies enable new production routes that could lift production-induced constraints on otherwise demonstrated radioisotopes like these. A proposed production route for a radioisotope matters clinically only if it addresses a bottleneck that is highly specific to each radioisotope.

## The Emerging Frontier

Many radioisotopes are known to be well-matched to a clinical goal, but their clinical evidence has not yet matured. Typically, this is because routine production or radiochemistry has not been developed—the emerging radioisotopes are not made in large enough quantities to be studied at clinical scales—or because clinical trials are too preliminary. Frequently, these form a negative cycle; clinical trials are hampered by immature development, and impetus for further development is restrained by immature clinical data.

$^{225}$Ac and $^{211}$At are promising for targeted α therapy (TAT) but are constrained by scarcity of parent materials or limited access to a specialised particle beam. $^{67}$Cu and $^{47}$Sc are promising for $\beta^-$ therapy and theranostic uses but limited by low yields and contaminants inherent in the cross sections of many production routes.

### Actinium-225

$^{225}$Ac is an investigational radioisotope for TAT. Over its full decay chain, $^{225}$Ac emits four α particles and two $\beta^-$ particles with a 9·9 d half-life. The α particles deposit approximately 28 MeV over ranges below 100 μm; once targeted to cancer cells, this can provide intense, spatially confined irradiation of malignant tissue. Photons emitted by two of the short-lived daughters in its decay chain, $^{221}$Fr and

$^{213}Bi$, permit SPECT imaging and dosimetry. $^{225}Ac$'s most mature application is PSMA-targeted treatment of metastatic castration-resistant prostate cancer.[40] $^{225}Ac$ conjugates are also being investigated in neuroendocrine tumours and haematological malignancies, including acute myeloid leukaemia.

Current $^{225}Ac$ supply cannot support large-scale clinical use. Traditional $^{225}Ac$ supply derives from limited stocks of $^{229}Th$ generators. Thus, evidence remains dominated by early-phase and non-randomized studies, and no $^{225}Ac$ radiopharmaceutical is approved.

### Astatine-211

$^{211}At$ is a unique potential TAT radioisotope because of its single α (5·87 MeV, 42%, or 7·45 MeV after electron capture decay to $^{211}Po$, 58%): if the radioisotope breaks its chemical bonds when recoiling during α emission, there are no untargeted α-particles that follow. Its short half-life of 7·2 h also decreases dose delivered to healthy tissue.[26] The halogen properties of $^{211}At$ let it form covalent bonds, of interest for developing lipophilic $^{211}At$-labelled compounds.

Supply and chemistry challenges have limited investigation of $^{211}At$ to an even greater extent than $^{225}Ac$. $^{211}At$ is primarily produced by irradiating $^{209}Bi$ with α particles, but few cyclotrons accelerate α particles at sufficient energy and current for clinically relevant production.[37] Even after the radioisotope has been produced, strategies to label radiopharmaceuticals stably with $^{211}At$ remain under investigation.[38] Nonetheless, three early-phase trials provide proof-of-principle of TAT with $^{211}At$ for several cancers. Treatment with $^{211}At$ was well-tolerated and showed preliminary efficacy.[36,39]

### Copper-67

$^{67}Cu$ is an investigative therapeutic radioisotope that decays primarily by emitting $\beta^-$ particles with energies (mean 0·145 MeV, maximum 0·562 MeV) comparable to established therapeutic radioisotope $^{177}Lu$. Compared to $^{177}Lu$, its shorter half-life (61·8 h)[17] could shorten treatment time for patients receiving $\beta^-$ therapy. $^{67}Cu$ is interesting as a theranostic radioisotope because its accompanying γ-rays (91 keV, 7%; 93 keV, 16%; 184 keV, 49%) are suitable for SPECT,[24] and pre-treatment PET can be facilitated via theranostic pairing with $^{64}Cu$.

$^{67}Cu$ was studied to treat non-Hodgkin lymphoma in a 1999 clinical trial; and is currently in a phase III trial for prostate cancer;[30] however, limited availability has obstructed most investigation of $^{67}Cu$. Both reactor-based and many accelerator-based routes have been investigated, but until recently, studied routes have been plagued by low yield and low purity due to co-production of $^{64}Cu$.

### Scandium-47

Beta-emitter $^{47}Sc$ is another investigative radioisotope for $\beta^-$ therapy. Compared to $^{177}Lu$ it has lower energies (mean 0·162 MeV, maximum 0·6 MeV)[17] and a shorter half-life (3·35 d)[17]. Its γ-rays (159 keV)[17] are suitable for SPECT.[24] $^{47}Sc$ has two chemically matched diagnostic partners, $^{44}Sc$ and $^{43}Sc$. $^{47}Sc$ production has come under investigation recently, and routes using reactors, cyclotrons, and e-linacs have been explored. Unfortunately, co-production of contaminants $^{48}Sc$ and $^{46}Sc$ makes many production routes not ideal for radiopharmaceutical purposes, and routes with greater purity tend to have low yields. Because a scalable production route for $^{47}Sc$ of clinical purity has not yet clearly emerged, preclinical studies have shown $^{47}Sc$'s potential,[24] but no clinical studies have been completed to date.

These examples demonstrate situations in which limited supply does more than restrict treatment volume. It restricts the accumulation of evidence needed to justify additional investment in the first place. Even so, significant research is under way, primarily in publicly-funded research institutions and in academia, to identify scalable production routes and develop corresponding methods and technologies to bring emerging radioisotopes into clinical use. Accelerator-based radioisotope production technologies have developed significantly in recent years and are responsible for many of the most promising production routes for these production-limited radioisotopes.

Even after a route suitable for clinical production has been identified, uncertain demand discourages private investment that can expand the production capacity needed to generate clinical evidence. This circularity is especially acute for radioisotopes requiring uncommon targets, beams, or separation systems, including the radioisotopes discussed above. Because of this lag in investment, focused development of physics technologies that are comparatively mature can be particularly important for emerging radioisotopes.

## Identifying Opportunities

The radioisotopes discussed above highlight three considerations that determine the potential value of an intervention in radioisotope production: the severity of the present constraint, the readiness of a technology that addresses it, and the clinical consequence if that intervention succeeds.

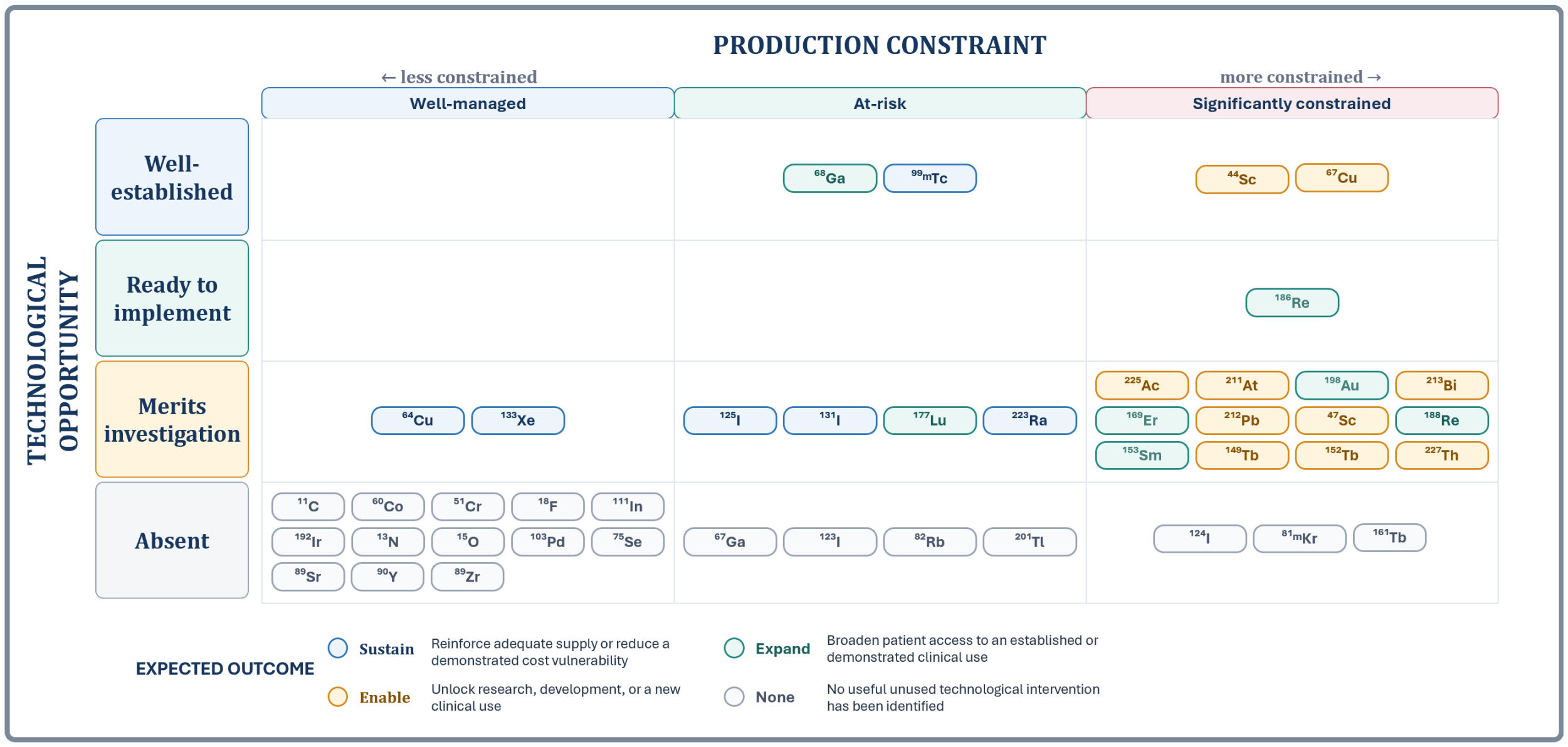

Figure 4. A potential technological intervention can have several beneficial effects on the supply of a radioisotope, especially where more mature technologies overlap with constrained production.

Figure 4 illustrates these potentials by category. The maturity of a technological intervention ranges from "absent" to "well-established." "Absent" refers to situations where no distinct, unused technological intervention was identified. "Merits investigation" technologies have theoretically mature physics and engineering, but important questions about efficacy or implementation remain.

"Ready-to-implement" refers to theoretically mature technologies that still require construction and characterization. "Well-established" technologies have been constructed and characterised.

Production constraints range from "well-managed" to "significantly constrained". "Well-managed" situations have existing production and distribution channels that are mature and serving the target patients as intended. This does not mean there is no room for improvement; an intervention may sustain supply by increasing robustness or decreasing cost. "At-risk" production may be adequate under ideal circumstances but has had a major regional, national, or widespread disruption that hampered patient access. "Significantly constrained" refers to radioisotopes with demonstrably inadequate production to support research and clinical use. Supply is also considered significantly constrained if it relies on sources that are not being replenished, like legacy stockpiles, or on production routes that cannot yield sufficiently high specific activity to permit clinically promising applications.

We categorise the expected benefit of a potential technological intervention as "none", "enable", "expand", or "sustain." "None" means no distinct, unused technological intervention with an expected useful benefit was identified. A radioisotope that could be "sustained" by an intervention is already available to patients but may benefit from improved supply robustness or decreased cost. A radioisotope that can be "expanded" currently has situations where it is unavailable when clinically desirable; the intervention would expand the patient base to which it is available. A radioisotope that would be "enabled" currently has supply-driven bottlenecks in research or treatment that the intervention could alleviate.

Several insights emerge from this comparison. (See Appendix B for a detailed breakdown of relevant production reactions.)

## Sustaining and expanding access to workhorse radioisotopes

Mature and quickly maturing accelerators can open alternative production paths to strengthen existing supplies of radioisotopes that are at-risk or have limited access in some regions (the intersection of "at-risk" and "sustain" or "expand" in Figure 4). Examples include direct cyclotron production of $^{68}$Ga and $^{99m}$Tc and production of $^{177}$Lu using a deuteron beam to augment reactor-based production.[42] However, these routes require careful expensive target materials, complex chemical processing, and careful management of contaminating radioisotopes.[2,42] Accelerator-driven thermal neutrons also merit investigation to augment reactor-based production of $^{99}$Mo ($^{99m}$Tc parent), $^{227}$Ac ($^{223}$Ra parent), $^{125}$I, $^{131}$I, and $^{133}$Xe. Accelerators, in a way that huge and expensive reactors cannot, could create regional production networks which could both expand supply and increase its resilience to disruption.

For radioisotopes already produced with accelerators, we recommend investment and investigation in high-current beams, high-power targets, automated processing, and enriched target recovery. For instance, higher-power proton cyclotrons, beam splitting across multiple characterised targets, and improved target cooling could sustain $^{64}$Cu production.

## Expanding applications of specific-activity-limited radioisotopes

Another priority is the production of radioisotopes with higher specific activity than is obtainable by direct neutron capture in reactors (the intersection of "expand" and "significantly constrained" in Figure 4), changing their clinically usable form. Charged-particle and photonuclear reactions producing $^{198}$Au,[43] $^{186}$Re, and $^{153}$Sm could provide no-carrier-added products for targeted radiopharmaceuticals beyond the bone and joint applications supported by carrier-added $^{186}$Re and

$^{153}Sm$. Accelerator-driven fast-neutron production through $^{169}Tm(n,p)^{169}Er$ and electromagnetic separation of reactor-produced $^{169}Er$ likewise merit investigation.[44] For $^{188}Re$, photonuclear production merits investigation as an alternative to production in the extremely rare reactors capable of producing $^{188}W$ generator feedstock.

## Enabling therapies bottlenecked by radioisotope scarcity

Modern high-power accelerators open several previously yield-limited pathways (the intersection of "enable" and "significantly constrained" in Figure 4). For $^{225}Ac$, photonuclear or accelerator-driven fast-neutron reactions could avoid the $^{227}Ac$ impurity inherent in proton spallation of $^{232}Th$. These interventions would also increase $^{225}Ac/^{213}Bi$ generator supplies. For $^{211}At$, irradiating bismuth targets with α particles remains the most practical, and α beams of sufficient energy and current are slowly growing in number. Higher-current α beams and advanced cooling strategies for the bismuth target could scale up production via this route.

Accelerator technologies, highly enriched target materials, and the necessary separation and recovery chemistry may overcome the challenges of $^{67}Cu$ and $^{47}Sc$ production. For $^{67}Cu$, photonuclear $^{68}Zn(\gamma,p)$, high-energy $^{68}Zn(p,2p)$, and $^{70}Zn(d,x)$ production should be developed in parallel until relative advantages in yield, purity, and target recovery are well-characterised.[45] For $^{44}Sc$, direct $^{44}Ca(p,n)^{44}Sc$ production has reached clinical use,[46] while higher-current proton production of $^{44}Ti$ could expand $^{44}Ti/^{44}Sc$ generator feedstock if the generator chemistry is further developed.[47] For $^{47}Sc$, indirect production through $^{46}Ca(n,\gamma)^{47}Ca \rightarrow ^{47}Sc$ or $^{48}Ca(\gamma,n)^{47}Ca \rightarrow ^{47}Sc$, ensuring radionuclidic purity, should be compared with direct proton production through proton irradiation and photonuclear production on $^{nat}V$, focusing on manageability of contaminants in the direct routes.

For $^{149}Tb$, proton spallation of $^{165}Ho$ or natural tantalum followed by on-line mass separation and heavy-ion production both merit investigation. For $^{152}Tb$, proton spallation of natural tantalum, high-energy proton production, and $^{12}C$-ion production merit investigation. For $^{212}Pb$, accelerator-driven fast neutrons could access $^{226}Ra(n,3n)^{224}Ra \rightarrow ^{212}Pb$, while high-energy proton production could expand $^{228}Th$ parent inventories. High-energy proton spallation of natural thorium could likewise expand $^{227}Ac$ and $^{227}Th$ inventories for $^{227}Th$ production.

Of note, several accelerator-based production routes for significantly constrained radioisotopes are well-established or ready-to-implement (see the intersection of these categories in Figure 4). These are the production routes we recommend to potential commercial suppliers for these radioisotopes. Examples from this analysis are photonuclear production of $^{67}Cu$ via $^{68}Zn(\gamma,p)^{67}Cu$,[40] production of $^{44}Sc$ using proton beams via $^{44}Ca(p,n)^{44}Sc$,[46] and no-carrier-added production of $^{186}Re$ via $^{186}W(p,n)^{186}Re$ or $^{186}W(d,2n)^{186}Re$.

These recommendations identify faculties of modern beam sources. Target designs capable of sustained high-power irradiation and recycling are an additional engineering challenge. For an opportunity classified as meriting investigation, the near-term goal should be a demonstration of integrated irradiation and processing that establishes yields and purities before more aggressive deployment studies.

# Conclusions and Outlook

Radioisotopes are invaluable to modern medicine, but their availability to patients relies on the successful merging of several fields. Clinical value emerges only when decay properties, radiochemistry, production, processing, distribution, and clinical workflow function as an integrated system. We reviewed over 40 current and emerging radioisotopes and categorised situations in which distinct, presently unused production technologies have the potential to address patterns of weakness in a complex supply landscape.

The appropriate intervention depends on the constraint. For well-managed radioisotopes, an alternative route warrants attention only where we identify a credible improvement in resilience or cost; technical feasibility alone is insufficient. For at-risk supplies, complementary regional production can reduce dependence on centralised facilities and vulnerable distribution chains. For significantly constrained radioisotopes, the priority is different: production advances may expand the amount available, provide the high-specific-activity material required, replenish long-lived generator parents, or make sustained clinical research possible for the first time.

Accelerator-driven neutron production, high-power photonuclear reactions, alternative parent inventories, and routes to no-carrier-added products are less mature technological opportunities that merit focused investigation. Nuclear reaction energetics and calculated yields for these approaches are maturing. The next step is integrated demonstration at clinically relevant scale, establishing target engineering and chemical processing, and validating yield and purity estimates. Several other technological opportunities are sufficiently mature to justify implementation or commercial evaluation, including established charged-particle routes for supply-limited therapeutic products and regional accelerator production of selected imaging radioisotopes. What could emerge from implementation of these technologies is an interlocking production network: centralised reactors and generator systems where they remain effective, regional accelerators where locality is critical, and specialised high-power facilities for radioisotopes requiring uncommon beams or complex separations. Prioritising technologies according to the severity of the supply constraint, the maturity of the complete production system, and the expected clinical consequence provides a practical basis for investment.

In the second paper of this Series, we present a detailed review of the status of reactors and accelerator-based production centres, the supply chain, its most prominent past disruptions, and established mitigation strategies to understand the opportunities and constraints associated with such an interlocking production network. We weigh the economics and regulation of distributed, accelerator-based regional production, asking how such a network would complement existing production and improve the reliability of established radiopharmaceuticals and access to emerging ones worldwide, including LMICs. The central opportunity is to convert advances in accelerator and target technology into reliable access to the right radioisotope, and ultimately to expand patient access and improve patient outcomes.

**Panel: Key Takeaways**

- Medical value depends on the entire pathway from isotope production and processing to distribution and clinical delivery.
- Supply constraints differ established radioisotopes need resilience, under-leveraged radioisotopes need enabling product forms or supply architectures, and emerging radioisotopes need sufficient availability.
- Technical interventions should address a defined bottleneck and provide a meaningful clinical benefit.
- Regional accelerators, alternative generator inventories, high-specific-activity products, and select photonuclear or accelerator-driven neutron routes offer distinct opportunities at different stages of maturity.
- The goal is a complementary production network that delivers the right radioisotope reliably and expands patient access.

## Declaration of interests

DW, JMC, JM, and DM report institutional research funding or salary support, all paid to MIT, from US Department of Energy awards DE-SC0024914 (which funded this work; see Acknowledgments) and DE-SC0024138, US National Science Foundation awards, a subaward under ARPA-E award DE-AR0002085 via Stony Brook University (DW, JM), and Heising-Simons Foundation grants (DW, JMC). They are developers of the HCHC-XX cyclotron concept described in this Review; MIT holds an MIT-funded contract with Ion Beam Applications SA (IBA) for its concept and feasibility phase, a no-cost agreement for testing at IBA is in preparation, and IBA has made no payments to the authors and holds no rights to the design. CH reports a US NSF Graduate Research Fellowship (DGE-2141064) and MIT-funded research visits to IBA as part of the MIT–IBA studies. PH reports institutional research funding from Siemens Healthineers, Radionetics, and Lilly, and is a co-investigator on Novartis-sponsored clinical trials. PH also reports participation on advisory boards for Telix Pharmaceuticals, Novartis, and BlueEarth Diagnostics, travel support from Telix Pharmaceuticals paid to the author, a signed consulting agreement with ZeEST Bio under which no consultation has yet taken place, institutional receipt of cyclotron facility support from GE Healthcare, and co-founder equity and a board seat in Epitorix, all outside the submitted work. SAE reports institutional research funding from the US National Institutes of Health (K08CA259626), Telix, Sofie, Novartis, and Lilly, and personal consulting fees from Telix, all outside the submitted work. All other authors declare no competing interests.

## Contributors

DW conceived the Series, developed the framework and search strategy, obtained funding, and led the project. JM and CH contributed equally to the literature search, data curation, analysis, figures and tables, and the original draft. DW drafted the framework and conclusions and revised the manuscript. PH, SAE, HS, FF, DM, and JRA contributed to conception and critically reviewed the clinical, imaging, and production content. JMC contributed to conception, obtained funding, and supervised the work with DW and PH. All authors critically revised the manuscript, had full access to all material, approved the final version, and accept responsibility for the decision to submit.

## Acknowledgments

This work was supported by the U.S. Department of Energy, Office of Science, Office of Isotope R&D and Production, under award number DE-SC0024914 (Principal Investigator DW, Co-Principal Investigator JMC). CH was supported by a U.S National Science Foundation Graduate Research Fellowship under grant number DGE-2141064.

The funders had no role in the design of the review, the literature search, the interpretation of the findings, the writing of the manuscript, or the decision to submit for publication. Ion Beam Applications SA had no role in the design, analysis, interpretation, or writing of this review, or in the decision to submit it. All authors had full access to all material reviewed and accept responsibility for the decision to submit for publication. Any opinions, findings, and conclusions or recommendations expressed in this material are those of the authors and do not necessarily reflect the views of the U.S. Department of Energy or the National Science Foundation.

## Data sharing

No new data were generated for this review. All information is drawn from published sources cited in the text and the appendix.

## Declaration of generative AI and AI-assisted technologies in the manuscript preparation process

During the preparation of this work the authors used Claude Opus 5, Claude Fable 5, and ChatGPT Sol 5.6 (2026) to check spelling and grammar, to comment on the readability and flow of draft sections, to assist with formatting the manuscript to journal requirements, and to support project organization. No scientific content, text, figures, tables, or references were generated by AI. After using these tool/service, the authors reviewed and edited the content as needed and take full responsibility for the content of the published article.

# References

1 Qaim S. Medical radioisotope production: science and technology. 2020. DOI:10.1515/9783110604375.

2 International Atomic Energy Agency. Alternative radioisotope production with a cyclotron. 2021. https://www-pub.iaea.org/MTCD/Publications/PDF/P1937_web.pdf.

3 Nuclear Energy Agency. Current trends in the supply and utilisation of medical radioisotopes. 2025. https://doi.org/10.82155/x9kb-7j83.

4 Lapi SE, Scott PJH, Scott AM, *et al.* Recent advances and impending challenges for the radiopharmaceutical sciences in oncology. *Lancet Oncol* 2024; **25**: e236–49.

5 Scott AM, Zeglis BM, Lapi SE, *et al.* Trends in Nuclear Medicine and the Radiopharmaceutical Sciences in Oncology: Workforce Challenges and Training in the Age of Theranostics. *Lancet Oncol* 2024; **25**: e250–9.

6 Giammarile F, Paez D, Zimmermann R, *et al.* Production and regulatory issues for theranostics. *Lancet Oncol* 2024; **25**: e260–9.

7 Abdel-Wahab M, Coleman CN, Eriksen JG, *et al.* Addressing challenges in low-income and middle-income countries through novel radiotherapy research opportunities. *Lancet Oncol* 2024; **25**: e270–80.

8 Stokke C, Kvassheim M, Blakkisrud J. Radioisotopes for targeted therapy: physical properties. *Molecules* 2022; **27**: 5429.

9 Wessels B, Meares C. Physical and chemical properties of radioisotope therapy. *Semin Radiat Oncol* 2000; **10**: 115–22.

10 Qaim S. New directions in nuclear data research for accelerator-based production of medical radioisotopes. *J Radioanal Nucl Chem* 2024; **333**: 3577–84.

11 International Atomic Energy Agency. Production and quality control of 225Ac radiopharmaceuticals. 2024.

12 Brink A, Cerci JJ, Giammarile F, *et al.* Global Challenges in the Provision of Theranostic Services the International Atomic Energy Agency (IAEA) Perspective. *World J Nucl Med* 2025; **24**: 238–43.

13 Novartis Annual Report 2025. Novartis. https://www.novartis.com/news/media-library/novartis-annual-report-2025 (accessed Sept 1, 2026).

14 Howell JN, Wilkes J, Maughan NM, *et al.* Industry Sponsorship and Disease Site Focus in Therapeutic Radiopharmaceutical Clinical Trials Over the Past Decade. *Int J Radiat Oncol Biol Phys* 2026; **124**: 575–80.

15 IAEA. Central Nervous System Radiotracer Development: Bench to Bedside, 1st ed. Vienna: International Atomic Energy Agency, 2024.

16 Radchenko V, Morgenstern A, Jalilian A, *et al.* Production and supply of alpha-particle-emitting radioisotopes for targeted alpha-therapy. *J Nucl Med* 2021; **62**: 1495–503.

17 National Nuclear Data Center. NuDat 3.0 database. https://www.nndc.bnl.gov/nudat3/.

18 International Atomic Energy Agency. Production and quality control of 18F labelled radiopharmaceuticals. 2021.

19 Boellaard R, Delgado-Bolton R, Oyen W, *et al.* FDG PET/CT: EANM procedure guidelines for tumour imaging: version 2.0. *Eur J Nucl Med Mol Imaging* 2015; **42**: 328–54.

20 US Food and Drug Administration. PET diagnostic radiopharmaceutical drugs: Medical Imaging Drugs Advisory Committee briefing document. 2023.

21 GE Healthcare Inc. Flyrcado (flurpiridaz F 18) injection: prescribing information. Arlington Heights (IL): GE Healthcare Inc, 2025 https://dailymed.nlm.nih.gov/dailymed/lookup.cfm?setid=3bb6caf8-ac98-46c7-a9f0-961e66032705.

22 Advanced Accelerator Applications USA Inc. Lutathera (lutetium Lu 177 dotatate) injection: prescribing information. Millburn (NJ): Advanced Accelerator Applications USA Inc, 2024 https://dailymed.nlm.nih.gov/dailymed/drugInfo.cfm?setid=72d1a024-00b7-418a-b36e-b2cb48f2ab55.

23 Banerjee S, Pillai MRA, Knapp FF (Russ). Lutetium-177 Therapeutic Radiopharmaceuticals: Linking Chemistry, Radiochemistry, and Practical Applications. *Chem Rev* 2015; **115**: 2934–74.

24 Mikolajczak R, Van Der Meulen NP, Lapi SE. Radiometals for imaging and theranostics, current production, and future perspectives. *Labelled Comp Radiopharmac* 2019; **62**: 615–34.

25 Uccelli L, Martini P, Pasquali M, Boschi A. Rhenium radioisotopes for medicine: production and applications. *Molecules* 2022; **27**: 5283.

26 Kampen W, Boddenberg-Pätzold B, Fischer M, *et al.* The EANM guideline for radiosynoviorthesis. *Eur J Nucl Med Mol Imaging* 2022; **49**: 681–708.

27 Quirijnen J, Han S, Zonnenberg B, others. Efficacy of 186Re-etidronate in prostate cancer patients with metastatic bone pain. *J Nucl Med* 1996; **37**: 1511–5.

28 Shinto A, Mallia M, Kameswaran M, others. Clinical utility of 188Re-hydroxyethylidene-1,1-diphosphonate as a bone pain palliative in multiple malignancies. *World J Nucl Med* 2018; **17**: 228–35.

29 Bernal P, Raoul J, Vidmar G, *et al.* Intra-arterial rhenium-188 lipiodol in the treatment of inoperable hepatocellular carcinoma: results of an IAEA-sponsored multinational study. *Int J Radiat Oncol Biol Phys* 2007; **69**: 1448–55.

30 Cardaci G, Baxi S, Vohra S, *et al.* Efficacy, safety, and patient reported outcomes of rhenium-skin cancer therapy for non-melanoma skin cancer: 1-year results from the EPIC-Skin study. *Adv Radiat Oncol* 2025; **10**: 101802.

31 International Atomic Energy Agency. Therapeutic radionuclide generators: 90Sr/90Y and 188W/188Re generators. Vienna: International Atomic Energy Agency, 2009 https://www.iaea.org/publications/8045/therapeutic-radionuclide-generators-90sr90y-and-188w188re-generators.

32 Ersöz O, Spink R, Griswold J, Yurt F, Mirzadeh S. Measurement of neutron capture cross section of 187W for production of 188W. *Appl Radiat Isot* 2019; **148**: 191–6.

33 Lantheus Medical Imaging Inc. Quadramet (samarium Sm 153 lexidronam) injection: prescribing information. 2017.

34 Vermeulen K, Van de Voorde M, Segers C, *et al.* Exploring the potential of high-molar-activity samarium-153 for targeted radionuclide therapy with [153Sm]Sm-DOTA-TATE. *Pharmaceutics* 2022; **14**: 2566.

35 Kratochwil C, Bruchertseifer F, Rathke H, *et al.* Targeted alpha-therapy of metastatic castration-resistant prostate cancer with 225Ac-PSMA-617: dosimetry estimate and empiric dose finding. *J Nucl Med* 2017; **58**: 1624–31.

36 Albertsson P, Bäck T, Bergmark K, *et al.* Astatine-211 based radionuclide therapy: Current clinical trial landscape. *Front Med* 2023; **9**: 1076210.

37 Feng Y, Zalutsky MR. Production, purification and availability of 211At: Near term steps towards global access. *Nuclear Medicine and Biology* 2021; **100–101**: 12–23.

38 Müller M, Pedersen NB, Shalgunov V, Jensen AI, Battisti UM, Herth MM. Astatine-211—Towards In Vivo Stable Astatine-211 Labeled Radiopharmaceuticals and Their (Pre)Clinical Applications. *Medicinal Research Reviews* 2026; **46**: 203–37.

39 Watabe T, Hatano K, Naka S, others. First-in-human SPECT/CT imaging of [211At]PSMA-5: targeted alpha therapy in a patient with refractory prostate cancer. *Eur J Nucl Med Mol Imaging* 2025; **52**: 2253–5.

40 Zimmermann R. Are $^{61}$ Cu,$^{64}$ Cu, and$^{67}$ Cu Really Happening? *J Nucl Med* 2026; **67**: 1032–5.

41 Ahmedova A, Todorov B, Burdzhiev N, Goze C. Copper radiopharmaceuticals for theranostic applications. *European Journal of Medicinal Chemistry* 2018; **157**: 1406–25.

42 Nagai Y, Kawabata M, Hashimoto S, *et al.* Estimated isotopic compositions of Yb in enriched 176Yb for producing 177Lu with high radionuclide purity by 176Yb(d,x)177Lu. *J Phys Soc Jpn* 2022; **91**: 044201.

43 Bhardwaj P, Claude G, Brenner W, Spreckelmeyer S. Radioactive gold isotopes in the era of theranostics: medical potential of 198Au and 199Au radioisotopes. *Nuklearmedizin* 2026; **65**: [published online ahead of print].

44 Talip Z, Borgna F, Müller C, *et al.* Production of mass-separated erbium-169 towards the first preclinical in vitro investigations. *Front Med (Lausanne)* 2021; **8**: 643175.

45 Nigron E, Guertin A, Haddad F, Sounalet T. Is 70Zn(d,x)67Cu the best way to produce 67Cu for medical applications? *Front Med (Lausanne)* 2021; **8**: 674617.

46 Singh A, Van Der Meulen NP, Müller C, *et al.* First-in-Human PET/CT Imaging of Metastatic Neuroendocrine Neoplasms with Cyclotron-Produced$^{44}$ Sc-DOTATOC: A Proof-of-Concept Study. *Cancer Biotherapy and Radiopharmaceuticals* 2017; **32**: 124–32.

47 Schmidt CE, Gajecki L, Deri MA, Sanders VA. Current State of 44Ti/44Sc Radionuclide Generator Systems and SeparationChemistry. *CRP* 2023; **16**: 95–106.

## Supplementary Material

# Contents

# Appendix A: Radioisotope Reference Table & Complete Radioisotope Overviews

| Isotope | Primar Decay Mode(s) | $t_{1/2}$ | Major Uses | Net Decay-Chain Outcome. (Mean Per Parent) | | | | |
|---|---|---|---|---|---|---|---|---|
| | | | | *End* | *α* | *β⁺* | *β⁻* | *γ* |
| $^{225}$Ac | α | 9·9 d | α-therapy | $^{209}$Bi | 4 | 0 | 2 | ✓ |
| $^{211}$At | EC (58%)<br>α (42%) | 7·21 h | α-therapy | $^{207}$Pb<br>$^{207}$Bi | 1 | 0 | 0 | ✓ |
| $^{213}$Bi | $β^-$ (98%)<br>α (2%) | 45·6 m | α-therapy | $^{209}$Bi | 1 | 0 | 2 | ✓ |
| $^{11}$C | $β^+$ | 20·3 m | PET metabolic & neurological, oncology/prostate applications | $^{11}$B | 0 | 1 | 0 | ✓ |
| $^{51}$Cr | EC | 27·7 d | RBC volume studies, GI bleeding, GFR estimation | $^{51}$V | 0 | 0 | 0 | ✓ |
| $^{60}$Co | $β^-$ | 5·3 y | Sterilization, external beam stereotactic surgery | $^{60}$Ni | 0 | 0 | 1 | ✓ |
| $^{64}$Cu | EC (44%)<br>$β^-$ (39%)<br>$β^+$ (17%) | 12·7 h | PET, prospective Cu-64/Cu-67 theranostic | $^{64}$Ni<br>$^{64}$Zn | 0 | 0·17 | 0·39 | ✓ |
| $^{67}$Cu | $β^-$ | 61·8 h | SPECT + β-therapy | $^{67}$Zn | 0 | 0 | 1 | ✓ |
| $^{169}$Er | $β^-$ | 9·4 d | Radiosynoviorthesis of small joints. | $^{169}$Tm | 0 | 0 | 1 | |
| $^{18}$F | $β^+$ (97%)<br>EC (3%) | 109·8 m | PET | $^{18}$O | 0 | 0·97 | 0 | ✓ |
| $^{67}$Ga | EC | 78·3 h | SPECT | $^{67}$Zn | 0 | 0 | 0 | ✓ |
| $^{68}$Ga | $β^+$ (89%)<br>EC (11%) | 67·7 m | PET (neuroendocrine & prostate) | $^{68}$Zn | 0 | 0·89 | 0 | ✓ |
| $^{198}$Au | $β^-$ | 2·7 d | Brachytherapy (LDR) | $^{198}$Hg | 0 | 0 | 1 | ✓ |
| $^{111}$In | EC | 2·8 d | SPECT (leukocyte, some neuroendocrine tumors) | $^{111}$Cd | 0 | 0 | 0 | ✓ |
| $^{123}$I | EC | 13·2 h | SPECT (thyroid & brain) | $^{123}$Te | 0 | 0 | 0 | ✓ |
| $^{124}$I | EC (77%)<br>$β^+$ (23%) | 4·2 d | PET (immune, thyroid) | $^{124}$Te | 0 | 0·23 | 0 | ✓ |
| $^{125}$I | EC | 59·4 d | Brachytherapy (LDR) | $^{125}$Te | 0 | 0 | 0 | ✓ |
| $^{131}$I | $β^-$ | 8·0 d | SPECT, β-therapy (hyperthyroidism, thyroid cancer) | $^{131}$Xe | 0 | 0 | 1 | ✓ |
| $^{192}$Ir | $β^-$ (95%)<br>EC (5%) | 73·8 d | Brachytherapy (HDR) | $^{192}$Pt<br>$^{192}$Os | 0 | 0 | 0·95 | ✓ |
| $^{81m}$Kr | Iso. Trans. (γ) | 13 s | Ventilation studies | $^{81}$Kr | 0 | 0 | 0 | ✓ |
| $^{212}$Pb | $β^-$ | 10·6 h | Prodrug for in-situ α-therapy | $^{208}$Pb | 1 | 0 | 2 | ✓ |
| $^{177}$Lu | $β^-$ | 6·65 d | SPECT + β-therapy | $^{177}$Hf | 0 | 0 | 1 | ✓ |
| $^{13}$N | $β^+$ | 10·0 m | PET (myocardial perfusion) | $^{13}$C | 0 | 1 | 0 | ✓ |
| $^{15}$O | $β^+$ | 2·04 m | PET (blood flow, perfusion measurements) | $^{15}$N | 0 | 1 | 0 | ✓ |
| $^{103}$Pd | EC | 17·0 d | Brachytherapy (LDR) | $^{103}$Rh | 0 | 0 | 0 | |
| $^{223}$Ra | α | 11·4 d | α-therapy | $^{207}$Pb | 4 | 0 | 2 | ✓ |
| $^{186}$Re | $β^-$ (93%)<br>EC (7%) | 3·7 d | SPECT+bone pain palliation, radiosynoviorthesis | $^{186}$Os<br>$^{186}$W | 0 | 0 | 0·93 | ✓ |
| $^{188}$Re | $β^-$ | 17·0 h | Bone pain palliation, skin cancer, hepatic cancer | $^{188}$Os | 0 | 0 | 1 | ✓ |
| $^{82}$Rb | $β^+$ (95%)<br>EC (5%) | 1·26 m | PET (myocardial perfusion) | $^{82}$Kr | 0 | 0 | 0·95 | ✓ |
| $^{153}$Sm | $β^-$ | 46·3 h | Bone pain palliation | $^{153}$Eu | 0 | 0 | 1 | ✓ |
| $^{44}$Sc | $β^+$ (94%)<br>EC (6%) | 4·0 h | PET (longer lived $^{68}$Ga sub.), 3γ imaging | $^{44}$Ca | 0 | 0·94 | 0 | ✓ |
| $^{47}$Sc | $β^-$ | 3·35 d | SPECT + targeted β-therapy | $^{47}$Ti | 0 | 0 | 1 | ✓ |
| $^{75}$Se | EC | 119·8 d | SeHCAT bile-acid malabsorption | $^{75}$As | 0 | 0 | 0 | ✓ |
| $^{89}$Sr | $β^-$ | 50·5 d | Bone pain palliation | $^{89}$Y | 0 | 0 | 1 | |
| $^{99m}$Tc | Iso. Trans. (γ) | 6·0 h | SPECT workhorse | $^{99}$Tc | 0 | 0 | 0 | ✓ |
| $^{149}$Tb | EC (76%)<br>α (17%)<br>$β^+$ (7%) | 4·1 h | PET + α-therapy | $^{149}$Sm<br>$^{145}$Nd | 0·17 | 0·07 | 0 | ✓ |
| $^{152}$Tb | EC (80%)<br>$β^+$ (20%) | 17·5 h | PET | $^{152}$Gd | 0 | 0·2 | 0 | ✓ |
| $^{161}$Tb | $β^-$ | 6·95 d | SPECT + β / Auger therapy | $^{161}$Dy | 0 | 0 | 1 | ✓ |
| $^{201}$Tl | EC | 72·9 h | SPECT myocardial perfusion; parathyroid localisation | $^{201}$Hg | 0 | 0 | 0 | ✓ |
| $^{227}$Th | α | 18·7 d | α-therapy | $^{207}$Pb | 5 | 0 | 2 | ✓ |
| $^{133}$Xe | $β^-$ | 5·3 d | Pulmonary ventilation/function imaging; cerebral blood-flow assessment | $^{133}$Cs | 0 | 0 | 1 | ✓ |
| $^{90}Y$ | $\beta^-$ | 64·0 h | Hepatic radioembolisation; targeted radioimmunotherapy | $^{90}Zr$ | 0 | 0 | 1 | |
| $^{89}Zr$ | EC (77%)<br>$\beta^+$ (23%) | 78·4 h | PET, particularly immuno-PET and imaging long-circulating biologics | $^{89}Y$ | 0 | 0·23 | 0 | ✓ |

Table A1: A summary of the key decay properties: mode(s), daughter products, half-lives, and major clinical uses for all radionuclides considered in this review.

# Established Clinical Core

## Carbon-11

### Radioisotope Overview

$^{11}C$ decays primarily via $\beta^+$ emission to $^{11}B$ with a half-life of 20·3 min and mean and maximum positron energy of 0·39 MeV and 0·96 MeV.[1] Carbon is ubiquitous in biomolecules, allowing it to incorporate into many endogenous compounds and drugs without modifying the biochemistry.

### Clinical Status

$^{11}C$ is an established but facility-limited PET isotope. Its short half-life permits serial same-day PET studies but requires rapid synthesis and use.[2] Principal clinical applications include glioma imaging and localization of recurrent prostate cancer, while research applications include imaging neurotransmission, amyloid deposition, metabolism, and drug pharmacokinetics.[2–4] Its 20-minute half-life largely confines use to centres capable of on-site production and radiochemistry. $^{11}C$ is diagnostic and has no canonical therapeutic partner.

### Production Mechanisms

$^{11}C$ is generally produced on site using an inexpensive nitrogen gas target and standard low-energy medical cyclotron.[2] We have identified no nuclear reaction pathways that offer a compelling alternative.

## Chromium-51

### Radioisotope Overview

$^{51}Cr$ decays by electron capture with a half-life of 27·7 days, emitting a single 320 keV gamma ~9·9% of the time with characteristic x-rays and auger electrons.[1]

### Clinical Status

The principal use for $^{51}Cr$ is as a red blood cell label. Chromate binds intracellular components of RBCs, allowing measurements of cell mass, survival, and blood volume. Splenic sequestration and occult bleeds can also be identified. $^{51}Cr$-EDTA is approved for measurements of kidney function.[5]

### Production Mechanisms

$^{51}Cr$ is reactor-produced by thermal-neutron capture on $^{50}Cr$ via the $^{50}Cr(n,\gamma)^{51}Cr$ reaction.[6] We have identified no nuclear reaction pathways that offer a compelling alternative.

## Cobalt-60

### Radioisotope Overview

$^{60}$Co decays by β– emission with a 5·27-year half-life and mean and maximum β energies of 0·10 MeV and 0·32 MeV. It decays into an excited state of $^{60}$Ni, with nearly every decay followed by a cascade of 1·173 MeV and 1·332 MeV gamma rays.[1]

### Clinical Status

$^{60}$Co is an established external-beam therapeutic isotope, especially where linear accelerators are cost-prohibitive. Multisource $^{60}$Co systems support stereotactic radiosurgery.[7] $^{60}$Co gamma irradiation is also widely used for bulk sterilisation of medical devices because it can process large, dense product volumes.[8]

### Production Mechanisms

$^{60}$Co is produced by irradiating stable $^{59}$Co in nuclear reactors.[6] Specific activity is primarily governed by neutron flux and irradiation time. This route is mature, uses inexpensive target material, and scales to very high activities. We have identified no nuclear reaction pathways that offer a compelling alternative.

## Copper-64

### Radioisotope Overview

$^{64}$Cu decays with a 12·7-h half-life by $\beta^-$ emission (38·5%; mean energy 0·191 MeV), $\beta^+$ emission (17·6%; mean energy 0·278 MeV), and electron capture (43·8%), with the resulting atomic relaxation producing Auger electrons.[1]

### Clinical Status

$^{64}$Cu is established for PET and PET/CT.[9] Its low-energy $\beta^+$ emission is similar to that of $^{18}$F and has a correspondingly short positron range, enabling high-resolution PET.[1,9] With $\beta^-$ and Auger electron decay products,[1] $^{64}$Cu could be therapeutic, but is more commonly discussed as a partner of therapeutic $^{67}$Cu.[10]

$^{64}$Cu's half-life is sufficient for labelling antibodies, antibody fragments, or peptides, and such agents have been studied clinically.[10]

### Production Mechanisms

Irradiation of $^{64}$Ni targets with protons or deuterons using a biomedical-energy cyclotron is standard for producing $^{64}$Cu at high yield and high purity.[11] The expensive $^{64}$Ni target may be recycled. Due to $^{64}$Ni's high melting point, there is potential to increase yield using higher-current beams.[11]

## Fluorine-18

### Radioisotope Overview

$^{18}F$ decays via positron emission (96·7%) and electron capture (3·3%) with a 109·8-min half-life. Its positrons have a mean and maximum energy of 0·25 MeV and 0·635 MeV.[1]

### Clinical Status

Fluorine's stable C-F chemistry lets it combine with a broad variety of tracer molecules.[1,12] An $^{18}F$-tagged glucose analogue, FDG, is the standard for imaging numerous cancers and metabolic disorders.[12,13] Additional approved uses include imaging bone turnover, amyloid or tau aggregation, myocardial perfusion, and many more.[14,15] Due to flexible chemistry, $^{18}F$ can bind the same ligand as a variety of therapeutic radioisotopes.

### Production Mechanisms

Routine $^{18}F$ production uses the $^{18}O(p,n)^{18}F$ reaction with highly enriched $H_2{}^{18}O$ using protons from compact medical cyclotrons. No-carrier-added aqueous $^{18}F$ can be separated and the enriched water recycled.[12] Automated nucleophilic labelling modules convert the fluoride into FDG or other compounds. This process is mature, scalable, and well-matched to clinical radiochemistry and regional distribution. We have identified no nuclear reaction pathways that offer a compelling alternative.

## Gallium-67

### Radioisotope Overview

$^{67}Ga$ decays by electron capture with a 3·26-day half-life, emitting principal photons at 93, 185, and 300 keV.[1,16]

### Clinical Status

$^{67}Ga$'s photon emissions permit delayed SPECT imaging after blood-pool clearance.

Administered as gallium citrate, $^{67}Ga$ binds transferrin and allows lesion localisation by reflecting binding and uptake resulting from inflammation or malignancies. Although increasingly displaced by PET, $^{67}Ga$ remains valuable for selected chronic infections, including vertebral osteomyelitis, particularly where PET is unavailable.[17,18]

### Production Mechanisms

$^{67}Ga$ is produced by irradiating $^{68}Zn$ with protons, yielding a no-carrier-added product. Production is mature. Especially in light of declining use, we have identified no nuclear reaction pathways that offer a compelling alternative.

## Gallium-68

### Radioisotope Overview

$^{68}$Ga decays via $\beta^+$ emission (89%) or electron capture (11%) to $^{68}$Zn with a 67·7 min half-life. Its positrons have mean and maximum energies of 0·89 MeV and 1·9 MeV.[1]

### Clinical Status

$^{68}$Ga is a PET tracer. Its half-life aligns with peptide pharmacokinetics, although its higher energy yields worse spatial resolution than $^{18}$F.[19] $^{68}$Ga-labelled somatostatin-receptor ligands, including DOTATATE and DOTATOC, are approved for PET imaging of neuroendocrine tumours. $^{68}$Ga–PSMA-11 is partnered with $^{177}$Lu for select prostate cancers.[19–21]

### Production Mechanisms

$^{68}$Ga is generally derived from $^{68}$Ge/$^{68}$Ga generators. However, activity per elution is limited, restricting high-throughput use. An alternative using a proton beam on $^{68}$Zn can provide larger batches with established accelerator technology.[22]

## Indium-111

### Radioisotope Overview

$^{111}$In decays by electron capture with a 2·80-day half-life, emitting 171 and 245 keV photons as well as low-energy Auger electrons.[1]

### Clinical Status

$^{111}$In has no established therapeutic role, but is a versatile radiolabel.[16] It can tag leukocytes for infection and inflammation studies.[23] It can also be used as a somatostatin-receptor tag for neuroendocrine tumours, though $^{68}$Ga and $^{64}$Cu are preferred.[21]

### Production Mechanisms

Irradiating cadmium with protons yields no-carrier-added $^{111}$In with a reusable target. Production and distribution are mature. Especially in light of declining use no compelling alternative production pathways are identified.

## Iodine-123

### Radioisotope Overview

$^{123}$I decays by electron capture, emitting single 159 keV $\gamma$-rays, with a 13·2 h half-life.[1]

### Clinical Status

$^{123}$I is an established SPECT isotope, with $\gamma$ energies near that of $^{99m}$Tc.[1]

Because the thyroid actively takes up iodine, $^{123}$I has been a longstanding isotope for thyroid imaging.[24] [$^{123}$I]-MIBG is FDA-approved for imaging some neuroendocrine tumours and to prognose heart failure.[24] [$^{123}$I]-ioflupane, targeting striatal dopamine transporters, is FDA-approved for brain imaging to diagnose Parkinsonian syndromes.

### Production Mechanisms

$^{123}I$ can be produced with high purity and at clinically relevant quantities using biomedical cyclotrons by bombarding $^{124}Xe$ with protons.[25] We have identified no compelling alternatives.

## Iodine-124

### Radioisotope Overview

$^{124}I$ decays by either $\beta^+$ emission, with mean energy 820 keV, or electron capture, resulting in Auger electrons and γ-rays.[1] It has a half-life of 4·18 d.[1]

### Clinical Status

Naturally targeted to thyroid, $^{124}I$ is used for thyroid cancer imaging and pretherapy dosimetry.

$^{124}I$'s half-life is suited to the biodistribution times of antibodies, making it attractive for immuno-PET. $^{124}I$-labelled antibodies cG250, huA33, and Trastuzumab have shown clinical promise for targeted PET of cancers.[24] PET using $^{124}I$ is complicated by the low frequency of its high-energy $\beta^+$ and its background γ spectrum.[24]

### Production Mechanisms

Biomedical proton cyclotrons produce $^{124}I$ at the highest obtainable purity by irradiating $^{124}Te$.[24] We have identified no compelling alternatives.

## Iodine-125

### Radioisotope Overview

$^{125}I$ decays by electron capture with a 59·4-day half-life, primarily producing low-energy 27–31 keV X-rays, a 35·5 keV gamma, and short-range conversion and Auger electrons.[1]

### Clinical Status

$^{125}I$ is principally used as a sealed brachytherapy source. The low-energy products attenuate rapidly, giving highly localised dose. The long half-life provides prolonged low-dose irradiation. Permanent seeds implanted in localised prostate cancer, alone or with external-beam radiotherapy, have proved effective.[26,27] Temporary episcleral plaques can also be applied to treat uveal melanomas.[28]

### Production Mechanisms

$^{125}I$ is produced in reactors through neutron capture on $^{124}Xe$, producing $^{125}Xe$ which decays to $^{125}I$. Gaseous separation yields high specific activity.[6,27] Production is mature and effective, but centralised. An accelerator driven neutron beam could augment and stabilise supply.

## Iodine-131

### Radioisotope Overview

$^{131}I$ decays by $\beta^-$ emission with an 8·02 d half-life and mean and maximum energy of 0·19 MeV and 0·61 MeV.[1] It co-emits γ-rays.[1]

### Clinical Status

$^{131}I$ has been a workhorse $\beta^-$ therapy isotope for decades. $^{131}I$ is naturally targeted to the thyroid, making it a standard treatment for hyperthyroidism and thyroid cancers. $^{131}I$-MIBG was FDA-approved to treat pheochromocytoma and paraganglioma (discontinued commercially) and has been studied in clinical trials for neuroblastoma.[24]

$^{131}I$ pharmaceuticals have also been approved for SPECT imaging,[24] though routine imaging use is not optimal because of high radiation dose.

### Production Mechanisms

$^{131}I$ is made in reactors. It can be produced by irradiating $^{130}Te$ and is also a major fission product of $^{235}U$.[29]

## Iridium-192

### Radioisotope Overview

$^{192}Ir$ has a 73·8-day half-life and decays primarily via $\beta^-$ emission to $^{192}Pt$, with a smaller electron-capture branch ending at $^{192}Os$. It has a complex photon spectrum extending from 0·1–1·4 MeV, with an effective energy near 0·35–0·40 MeV.[1,30]

### Clinical Status

The high specific activity of $^{192}Ir$ permits a small source to deliver high doses, making it an important radioisotope for high-dose-rate brachytherapy.[30] The photon energy provides therapeutic levels of tissue penetration, while allowing modest shielding. The 74-d half-life requires source replacement every few months, creating a recurring, but established, manufacturing and logistics requirement. $^{192}Ir$ is used as a sealed radiation source rather than as a molecular radiopharmaceutical and has no theranostic partner.

### Production Mechanisms

$^{192}Ir$ is produced by thermal-neutron capture on $^{191}Ir$ in reactors. The large capture cross-section supports high-activity sources.[6,31] Because product and target are isotopes of the same element, the material is carrier-added, but this does not limit sealed-source use. Proton or deuteron irradiation of enriched $^{192}Os$ can produce no-carrier-added $^{192}Ir$, but yields are low and enriched-target recovery is complex. These accelerator routes have not displaced mature reactor production.

## Krypton-81m

### Radioisotope Overview

$^{81m}$Kr decays via isomeric transition with a 13·1-s half-life, emitting a 190 keV photon. It is biologically inert.[1]

### Clinical Status

$^{81m}$Kr is used in planar and SPECT ventilation-perfusion imaging, primarily for pulmonary embolism and assessment of regional lung function. During continuous inhalation, it reaches an alveolar steady state allowing activity to reflect ventilation. Its ultrashort half-life permits rapidly repeated acquisitions at low patient dose.[16,32] Its 190-keV photon can be separated from $^{99m}$Tc's 140-keV photon in simultaneous ventilation/perfusion imaging.[32,33]

### Production Mechanisms

$^{81m}$Kr is obtained from an $^{81}$Rb/$^{81m}$Kr generator. $^{81}$Rb is produced using protons from a compact cyclotron on $^{82}$Kr. The target is recoverable. The parent's 4·6-h half-life restricts each generator to same-day distribution and makes low-throughput use expensive.

## Lutetium-177

### Radioisotope Overview

$^{177}$Lu decays by $\beta^-$ emission to $^{177}$Hf with a 6·65 d half-life and mean and maximum β energies of 0·134 MeV and 0·498 MeV.[1] It co-emits gamma-rays including 113 and 208-keV photons.[1]

### Clinical Status

$^{177}$Lu's moderate-energy $\beta^-$ creates cross-fire that deposits dose across a diameter of about 20 cells.[34] $^{177}$Lu has recently become a workhorse therapeutic radioisotope due to application in peptide receptor radionuclide therapy: $^{177}$Lu-based radiopharmaceuticals are FDA-approved to treat prostate cancer (Pluvicto) and neuroendocrine tumours (Lutathera). Its γ-rays permit post-treatment SPECT imaging and dosimetry.[35] $^{177}$Lu is often paired with $^{68}$Ga for PET.[34]

### Production Mechanisms

Irradiating enriched $^{176}$Lu in reactors produces $^{177}$Lu in large quantities, but the carrier-added product has specific activity too low for many treatments.[10] Irradiating $^{176}$Yb in medium- and high-flux reactors yields no-carrier-added $^{177}$Lu, though with a small cross-section.[10] Large-scale separation of $^{177}$Lu from ytterbium is complex and not yet standardised in academic literature.[10]

Reactor-based production routes are effective, but an accelerator-based route irradiating $^{176}$Yb with deuterons may be able to augment production.[36]

## Nitrogen-13

### Radioisotope Overview

$^{13}N$ decays almost entirely by positron emission with a 9·97-min half-life and mean and maximum positron energies of 0·492 MeV and 1·2 MeV. As a major component of many biomolecules, nitrogen has flexible administration chemistry.[1,16]

### Clinical Status

$^{13}N$ is an established tracer for myocardial perfusion, used for rest/stress PET imaging.[37] The short half-life enables dynamic imaging of absolute blood flow.[38]

### Production Mechanisms

$^{13}N$ is usually produced by irradiating water with low-energy protons. Resulting nitrogen species are rapidly converted to ammonia and purified.[16] Production is inexpensive and uses no enriched target. However, the short half-life requires on-site production and processing. This infrastructure constraint means limited adoption despite strong clinical performance.

## Oxygen-15

### Radioisotope Overview

$^{15}O$ decays almost exclusively via positron emission to $^{15}N$ with a 2·04 min half-life; its positrons have mean and maximum energies of 0·735 MeV and 1·7 MeV.[1]

### Clinical Status

$^{15}O$'s short half-life allows rapid serial measurements at low dose. It is typically administered in water where it diffuses freely into and out of blood, making it a standard for absolute perfusion measurements. It is primarily used in myocardial PET to quantify myocardial flow and flow reserve, regarded as the most accurate of non-invasive measurements. It can also quantify perfusion in brain and tumour tissue.[39]

### Production Mechanisms

$^{15}O$'s short half-life necessitates local acquisition and rapid use. Production generally uses a deuteron beam on $^{14}N$. An alternative uses a proton beam on enriched $^{15}N$.[40] Direct production with a water target using an electron beam has shown feasibility.[41]

## Palladium-103

### Radioisotope Overview

$^{103}Pd$ decays via electron capture with a half-life of 17 days. It emits low-energy X-rays and Auger electrons.[1]

### Clinical Status

The low energy of $^{103}Pd$ decay products makes the delivered dose highly localised. Because of this, $^{103}Pd$ is clinically established for low-dose-rate brachytherapy where it is implanted directly in a tumour, e.g. for prostate cancer.[42] The implant is permanent since $^{103}Pd$ decays away sufficiently quickly.

### Production Mechanisms

$^{103}$Pd is produced with good purity by irradiating $^{103}$Rh with protons or deuterons using biomedical energy cyclotrons.[29]

## Radium-223

### Radioisotope Overview

$^{223}$Ra decays with an 11·4-d half-life through a chain of short-lived daughters to stable $^{207}$Pb, collectively emitting four α particles along with lower-energy β⁻ particles and photons.[1]

### Clinical Status

The high LET and short range of the $^{223}$Ra decay-chain daughters, below 100 µm, allow intense and spatially restricted irradiation. Low-energy photons permit activity measurement and limited imaging.[43] Administered as $^{223}$Ra dichloride, $^{223}$Ra behaves as a calcium analogue and localises to areas of increased bone turnover. It is approved for castration-resistant prostate cancer with symptomatic bone metastases and no known visceral disease.[43,4443,43,44,44]

### Production Mechanisms

$^{223}$Ra is primarily obtained from $^{227}$Ac generators, which decay through short-lived intermediates to $^{223}$Ra. The 21·8-y $^{227}$Ac half-life makes centralised production and distribution of generators feasible and efficient.[45] The generator route remains the most mature and effective, but there is potential to augment centralised reactors with accelerator-driven neutron beams that could lend additional supply stability.

## Rubidium-82

### Radioisotope Overview

$^{82}$Rb decays to $^{82}$Kr with a 1·26 min half-life, primarily via positron emission with a mean and maximum energy of 1·5 MeV and 3·4 MeV.[1]

### Clinical Status

$^{82}$Rb is established for PET myocardial perfusion imaging. Administered as rubidium chloride it behaves as a potassium analogue and is preferentially retained by viable myocardium. Its short half-life allows low-dose, rapid sequence imaging in a single visit which can quantify absolute myocardial blood flow and flow reserve.[46] The relatively high energy yields lower spatial resolution than $^{13}$N or $^{18}$F.[47,48]

### Production Mechanisms

$^{82}$Rb is derived from an $^{82}$Sr/$^{82}$Rb generator. The $^{82}$Sr parent is derived from high energy protons on $^{85}$Rb. The $^{82}$Sr is chemically separated and loaded into generators to be distributed. $^{82}$Rb's short half-life makes a generator ideal. We have identified no nuclear reaction pathways that offer a compelling alternative.

## Selenium-75

### Radioisotope Overview

$^{75}$Se decays by electron capture to $^{75}$As with a 119·8-d half-life. It emits 136 and 265 keV photons.[1]

### Clinical Status

$^{75}$Se is used primarily as an orally administered bile-acid analogue in the SeHCAT test. The principal photons and longer half-life permit gamma-camera measurements for an extended time after administration and support diagnosis of bile-acid malabsorption.[49]

### Production Mechanisms

Commercial SeHCAT production uses neutron irradiation of enriched $^{74}$Se, followed by chemical incorporation into tauroselcholic acid.[50] An alternative uses protons from a small cyclotron on $^{75}$As. This yields high-purity $^{75}$Se; however, demand has not pushed this to displace the established reactor supply.[51]

## Strontium-89

### Radioisotope Overview

$^{89}$Sr decays by $\beta^-$ emission to $^{89}$Y with a 50·5-day half-life and mean and maximum β energies of 0·58 and 1·46 MeV.[1]

### Clinical Status

$^{89}$Sr is approved for palliation of bone lesions from prostate or breast metastases. As a calcium analogue it accumulates at sites of increased osteoblastic activity.[52] Its prolonged irradiation can provide durable relief, however, unlike $^{223}$Ra, it has no established survival benefit.

### Production Mechanisms

$^{89}$Sr is produced by neutron irradiation of enriched $^{88}$Sr in reactors.[53] A no-carrier-added product can be obtained by irradiating $^{89}$Y with fast neutrons. An accelerator-driven fast- or thermal-neutron source could alternatively drive these processes to improve supply.[54]

## Technetium-99m

### Radioisotope Overview

$^{99m}$Tc decays by isomeric transition to $^{99}$Tc with a 6·0-h half-life, emitting a 140·5-keV photon.[1]

### Clinical Status

$^{99m}$Tc's half-life permits radiopharmaceutical preparation, uptake, and imaging while limiting patient radiation exposure. Its photons are well suited to planar and SPECT imaging. $^{99m}$Tc is used in approximately 40 million procedures annually, representing about 80% of nuclear-medicine examinations. Major applications include bone, myocardial-perfusion, renal, hepatobiliary, and lung imaging.[55] $^{99m}$Tc-MAA is routinely used for pretreatment planning for $^{90}$Y radioembolisation, while receptor-targeted $^{99m}$Tc agents are being developed as companion diagnostics alongside therapeutic radioisotopes.[56,57]

### Production Mechanisms

Most $^{99m}$Tc is supplied through $^{99}$Mo/$^{99m}$Tc generators. $^{99}$Mo is conventionally obtained by thermal-neutron fission of $^{235}$U in reactors. It is chemically isolated and loaded into generators.[55] This system remains the most effective means of distributed, high-volume supply. The leading accelerator alternative is direct production through $^{100}$Mo(p,2n)$^{99m}$Tc using approximately 10–25 MeV protons.

This permits regional cyclotron production but requires enriched $^{100}Mo$, rapid chemical processing and target recycling, and careful control of co-produced technetium isotopes.[58]

## Thallium-201

### Radioisotope Overview

$^{201}Tl$ decays via electron capture to $^{201}Hg$ with a 72·9-h half-life. Its emissions are dominated by mercury x-rays from 69-83 keV, with additional 135- and 167-keV gamma rays suitable for planar and SPECT imaging.[1]

### Clinical Status

$^{201}Tl$ is approved for myocardial-perfusion imaging to identify reversible ischemia and irreversible infarction. When administered as thallous chloride, it behaves as a potassium analogue and is taken up by viable myocardium where it provides measurements of perfusion and cellular membrane integrity. Redistribution after injection can provide information on myocardial viability without a second tracer administration.[59] Its relatively low photon energies and longer half-life give less favourable image quality and dosimetry than $^{99m}Tc$ agents, so use has declined, but its redistribution behaviour remains clinically useful. $^{201}Tl$ is diagnostic and has no canonical therapeutic partner.

### Production Mechanisms

$^{201}Tl$ is produced with medium-energy protons on enriched $^{203}Tl$ via $^{203}Tl(p,3n)^{201}Pb$, followed by the 9·3-h decay of $^{201}Pb$ to $^{201}Tl$. Irradiation is typically performed over approximately 20–30 MeV, followed by cooling and chemical separation from the thallium target and lead intermediates. Energy control and cooling limit co-produced $^{200}Pb/^{200}Tl$ and $^{202m}Pb/^{202}Tl$.[16,31] This centralised cyclotron route is mature and serves current demand; no alternative production pathway offers a clear access advantage.

## Xenon-133

### Radioisotope Overview

$^{133}Xe$ decays via $\beta^-$ emission to $^{133}Cs$ with a 5·25-d half-life. Its beta spectrum has mean and maximum energies of approximately 0·10 and 0·35 MeV. An 81-keV photon is emitted in about 37% of decays.[1]

### Clinical Status

$^{133}Xe$ is an inert diagnostic gas administered by inhalation. Sequential measurements permit evaluation of pulmonary ventilation and air trapping. The gas has also been used to assess cerebral blood flow.[60] Its multi-day half-life supports centralised distribution. However, the low photon energy limits spatial resolution and requires a closed system to prevent room contamination. $^{99m}Tc$ aerosols and $^{81m}Kr$ provide alternatives, but $^{133}Xe$ remains an established option where ventilation equipment and supply are available. It has no therapeutic role or canonical theranostic partner.

### Production Mechanisms

Commercial $^{133}Xe$ is primarily obtained from $^{235}U$ fission. After irradiation and cooling, xenon is separated from other fission products and distributed as a carrier-gas mixture.[6] The 5·25-d half-life supports centralised reactor production and distribution. The established route is efficient, although it shares the waste-management and supply-concentration limitations of other fission products; no accelerator route currently offers a compelling practical alternative.

## Yttrium-90

### Radioisotope Overview

$^{90}Y$ decays via $\beta^-$ emission to $^{90}Zr$ with a 64·0-h half-life and mean and maximum β energies of approximately 0·93 MeV and 2·28 MeV.[1] It emits essentially no useful primary photons. Minor bremsstrahlung and minor internal-pair-production photons are emitted.

### Clinical Status

$^{90}Y$ is established for radioembolisation of selected primary and metastatic liver tumours.[61,62] $^{90}Y$-labelled ibritumomab tiuxetan is also approved for selected B-cell non-Hodgkin lymphomas. The high-energy beta emission is well-suited for irradiation of macroscopic disease, but requires careful treatment planning to minimise normal-tissue dosimetry.

### Production Mechanisms

No-carrier-added $^{90}Y$ is generally eluted from long-lived $^{90}Sr/^{90}Y$ generators. $^{90}Sr$, in turn, is recovered from $^{235}U$ fission products and, with its 28·8-year half-life, supports repeated generator use.[63,64] The fission-product generator route is mature and supplies current radiopharmaceutical use, with no demonstrated accelerator alternative that clearly improves purity or distribution.

## Zirconium-89

### Radioisotope Overview

$^{89}Zr$ decays via electron capture (77·3%) and $\beta^+$ emission (22·7%) to $^{89}Y$ with a 78·4-h half-life. The mean and maximum positron energies are about 0·40 and 0·90 MeV. A 909-keV photon accompanies nearly every decay.[1,65]

### Clinical Status

$^{89}Zr$ is an established immuno-PET platform. Its half-life matches the biodistribution of monoclonal antibodies and other large biomolecules, allowing imaging for days after administration. $^{89}Zr$-labelled antibodies have been used to assess target expression, whole-body biodistribution, and organ uptake. It has been demonstrated for planning and dosimetry in antibody-based therapies.[65,66] Individual agents remain largely investigational, and stable chelation is vital as released zirconium accumulates in bone. Its theranostic role generally comes from labelling the same biological vector used for therapy rather than from a canonical isotope pair.

### Production Mechanisms

$^{89}Zr$ is routinely produced by irradiating monoisotopic $^{89}Y$ with protons. The metallic yttrium target is inexpensive, and hydroxamate-based separation yields high-purity, no-carrier-added $^{89}Zr$.[65,67] The route is compatible with low-energy medical cyclotrons but requires solid-target irradiation and chemical processing. Its maturity and the 78-h distribution window make regional production practical; no alternative nuclear pathway offers a compelling advantage.

# Underleveraged Middle

## Erbium-169

### Radioisotope Overview

$^{169}$Er decays via $\beta^-$ emission with a 9·39-day half-life to $^{169}$Tm, with mean and maximum β energies of approximately 0·099 MeV and 0·34 MeV, respectively. Internal conversion contributes additional low-energy electrons.[1]

### Clinical Status

The $\beta^-$ and additional electrons have a small maximum range in tissue of about 1 mm.[68,69] $^{169}$Er is established to treat small joint inflammation, especially persistent, steroid-resistant rheumatoid arthritis. Due to low penetration of $^{169}$Er's radiation, and potential theranostic pairings with $^{68}$Ga or $^{44}$Sc, there has been recent interest in developing $^{169}$Er-based radiopharmaceuticals.[69]

### Production Mechanisms

$^{169}$Er is produced by irradiating highly enriched $^{168}$Er in reactors. The modest cross-section and stable inseparable target keep specific activity low. This is sufficient for radiosynoviorthesis [70] but potential radiopharmaceuticals require high specific activity.

Electromagnetic separation of reactor-produced $^{169}$Er can achieve this, but with low efficiency.[70] An accelerator-driven neutron beam on $^{169}$Tm could produce no-carrier-added $^{169}$Er.[31] Reactor-driven production is adequate for present needs, but accelerator-driven routes to high-specific-activity $^{169}$Er may open new therapeutic uses.

## Gold-198

### Radioisotope Overview

$^{198}$Au emits $\beta^-$ particles with a mean energy 0·313 MeV and 411 keV γ-rays with a 2·7 d half-life.[1]

### Clinical Status

$^{198}$Au is clinically established as a low-dose-rate brachytherapy isotope, e.g. for oral cancers.[42] Its half-life is also long enough to allow incorporation into radiopharmaceuticals, which could provide $\beta^-$ therapy of large or heterogeneous tumours and SPECT imaging.[71]

### Production Mechanisms

The established production route is irradiating $^{197}$Au in reactors,[71] but the carrier-added product has low specific activity. Deuteron-driven reactions on platinum and photonuclear reactions on mercury show promise but are hindered by difficult separation chemistry. Methods to separate gold from platinum or mercury have matured enough to explore $^{198}$Au radiopharmaceuticals requiring high specific activity.[71]

## Rhenium-186

### Radioisotope Overview

$^{186}Re$ decays predominantly by $\beta^-$ emission with a 3·72-day half-life and mean and maximum $\beta^-$ energies of 0·35 MeV and 1·08 MeV. A 137 keV photon is emitted 9·5% of the time.[1]

### Clinical Status

$^{186}Re$ sulfide is established for radiosynoviorthesis of medium-sized joints affected by persistent inflammatory synovitis.[68] It has also demonstrated palliation of bone pain from metastases.[72] The 137 keV photon emitted allows concurrent SPECT imaging.[73]

### Production Mechanisms

Production is generally done with a neutron beam on $^{185}Re$. This yields a carrier-added product suitable for bone and joint applications, but limits uses requiring high specific activity. An alternative could use protons or deuterons on $^{186}W$.[74]

## Rhenium-188

### Radioisotope Overview

$^{188}Re$ decays by $\beta^-$ emission with a 17·0-h half-life and mean and maximum β energies of 0·78 and 2·12-MeV. A 155 keV photon is emitted in approximately 15% of decays.[1]

### Clinical Status

$^{188}Re$–HEDP has demonstrated use for palliation of metastatic bone pain, while transarterial $^{188}Re$–lipiodol has shown efficacy against hepatocellular carcinoma.[75,76] Recent work on topical formulations has shown promise for basal and squamous-cell carcinoma.[77] The 155 keV photon permits concurrent SPECT imaging and dosimetry.

### Production Mechanisms

$^{188}Re$ is generally obtained from $^{188}W/^{188}Re$ generators. The $^{188}W$ parent is produced in reactors by successive neutron capture on enriched $^{186}W$. The second capture competes with the intermediate $^{187}W$ decay, so exceptionally high neutron flux is required, restricting production to a few reactors.[64,78] An alternative photonuclear reaction on $^{192}Os$ may provide a viable alternative.

## Samarium-153

### Radioisotope Overview

$^{153}Sm$ decays by $\beta^-$ emission with a 1·93-day half-life and mean and maximum β energies of 0·233 MeV and 0·803 MeV. A 103 keV photon is emitted in 29% of decays.[1]

### Clinical Status

$^{153}Sm$ is primarily used for palliative care in bone metastases. When administered as $^{153}Sm$-EDTMP it localises to calcium-phosphate compound sites in bone and relieves painful lesions. The $\beta^-$ dose is well localised, with a tissue range of 0·5–3 mm. The photon permits post-treatment SPECT imaging.[79]

High specific activity $^{153}Sm$ has demonstrated receptor-specific tumour uptake, SPECT visualisation, and cytotoxicity in preclinical studies.[80]

### Production Mechanisms

Routine production produces $^{153}Sm$ by irradiating $^{152}Sm$ in reactors. This allows high yields, but the product is carrier-added, which is suitable for EDTMP, but limits applications requiring higher specific activity. No-carrier added $^{153}Sm$ for such applications could be produced by irradiating $^{150}Nd$ with an α beam.[81]

# Emerging Frontier

## Actinium-225

### Radioisotope Overview

$^{225}Ac$ decays with a 9·92-day half-life through a chain of short-lived daughters to $^{209}Bi$, collectively emitting four α particles and two $\beta^-$ particles. The α particles deposit approximately 28 MeV per complete decay chain.[1]

### Clinical Status

$^{225}Ac$ is an investigational radioisotope for targeted α therapy (TAT). $^{225}Ac$ deposits significant energy over ranges below 100 µm, enabling intense, spatially confined irradiation. Photons emitted by $^{221}Fr$ and $^{213}Bi$ permit SPECT imaging and dosimetry. Its most mature application is PSMA-targeted treatment of metastatic castration-resistant prostate cancer.[82] $^{225}Ac$ conjugates are also being investigated in neuroendocrine tumours and hematological malignancies, including acute myeloid leukemia. However, evidence remains dominated by early-phase and non-randomized studies, and no $^{225}Ac$ radiopharmaceutical is currently approved.

### Production Mechanisms

Traditional $^{225}Ac$ supply derives from $^{229}Th$ generators. High-energy proton spallation of $^{232}Th$ provides larger yields but co-produces long-lived $^{227}Ac$.[83] The most promising alternative is photonuclear production of the parent isotope $^{225}Ra$, followed by $\beta^-$ decay to $^{225}Ac$; this generator route is scalable and avoids $^{227}Ac$ contamination.[84] $^{225}Ra$ can similarly be produced using accelerator-driven fast neutrons.[85,86] Both routes require radiotoxic $^{226}Ra$ targets and efficient target recovery, while the neutron pathway additionally requires a sufficiently intense fast-neutron source.

## Astatine-211

### Radioisotope Overview

$^{211}At$ emits a single α during its decay, with energies of 5·87 MeV (42%) or 7·45 MeV (after electron capture decay to $^{211}Po$, 58%) with a half-life of 7·21 h.[17] $^{211}At$ displays both metallic and halogen chemical properties.

### Clinical Status

$^{211}At$ is unique for TAT because of its single α: if the radioisotope breaks its chemical bonds when recoiling during α emission, there are no untargeted α-particles that follow. Its short half-life also decreases dose delivered to healthy tissue.[87]

The halogen properties of $^{211}At$ let it form covalent bonds, of interest for developing lipophilic $^{211}At$-labelled compounds.[88] Strategies to label radiopharmaceuticals stably with $^{211}At$ remain under investigation.[88]

Supply and chemistry challenges limit investigation of $^{211}At$. Three early-phase trials provide proof-of-principle of TAT with $^{211}At$ for several cancers. Treatment with $^{211}At$ was well-tolerated and showed preliminary efficacy.[87,89]

### Production Mechanisms

$^{211}At$ is primarily produced by irradiating $^{209}Bi$ with α particles, but few cyclotrons accelerate α particles at sufficient energy and current for clinically relevant production.[90] Cyclotrons with higher current, together with advanced cooling strategies for the Bi target,[90] could potentially scale up $^{211}At$ production via this route.

## Bismuth-213

### Radioisotope Overview

$^{213}Bi$ is a short-lived α emitter used in TAT. It has a half-life of 45·6 min and decays through several intermediates ending at 209Bi. The net result is one α particle and two $\beta^-$ particles. It also emits a 440 keV photon, which can support imaging measurements.

### Clinical Status

$^{213}Bi$ has been clinically investigated in hematologic malignancies and solid tumours, including acute myeloid leukaemia (AML), gliomas, neuroendocrine tumours, melanomas, and prostate cancer. Early studies demonstrated feasibility and signs of efficacy, especially for $^{213}Bi$-lintuzumab in AML and $^{213}Bi$-DOTA-substance P analogues in gliomas. However, broad clinical adoption has not occurred.

### Production Mechanisms

Because of its short half-life, $^{213}Bi$ is not distributed directly; rather, it is generally obtained in situ from $^{225}Ac/^{213}Bi$ generators. No significant alternative channels are known, making $^{213}Bi$ availability closely tied to $^{225}Ac$ production.

## Copper-67

### Radioisotope Overview

$^{67}Cu$ decays primarily by $\beta^-$ emission with accompanying γ-rays (91 keV, 7%; 93 keV, 16%; 184 keV, 49%) with a half-life of 61·8 h and a mean and maximum $\beta^-$ energy of 0·145 MeV and 0·562 MeV.[1,9]

### Clinical Status

$^{67}Cu$'s $\beta^-$ energy is comparable to $^{177}Lu$, and its shorter half-life could shorten treatment time for patients receiving $\beta^-$ therapy. Its γ-rays are suitable for SPECT,[10] and $^{64}Cu$ would be a chemically matched diagnostic partner.

$^{67}Cu$ was studied to treat non-Hodgkin lymphoma in a 1999 clinical trial[9] and is currently in a phase III trial for prostate cancer.[91] Limited availability has obstructed investigation of $^{67}Cu$.

### Production Mechanisms

$^{67}Cu$ production has been plagued by low yield and low purity due to co-production of $^{64}Cu$ in reactor-based and many accelerator-based routes. Photonuclear production on $^{68}Zn$ shows potential for large-scale production of high-purity $^{67}Cu$ and has been used to supply a clinical trial.[10,91] Irradiating $^{70}Zn$ with deuterons or $^{68}Zn$ with high-energy protons are also promising routes.[10,92]

## Lead-212

### Radioisotope Overview

$^{212}Pb$ decays by $\beta^-$ emission to $^{212}Bi$ with a 10·6-h half-life, with mean and maximum β energy of 0·177 MeV and 0·57 MeV. The ensuing daughters emit one α at 6·1 or 8·8 MeV. A 239 keV photon is emitted in 44% of decays.[1]

### Clinical Status

$^{212}Pb$ radiopharmaceuticals remain investigational but have been explored for a range of targets. The most mature application is in SSTR-targeted agents for neuroendocrine tumours.[93] Human studies have also evaluated agents targeting prostate cancer, melanoma, peritoneal malignancies, and other solid tumours.[94–96]

### Production Mechanisms

$^{212}Pb$ is produced through the $^{228}Th$ decay chain. $^{228}Th$ parent is commonly recovered from legacy $^{232}U$ inventories. The $^{224}Ra$ daughter is separated and loaded onto generators from which $^{212}Pb$ is eluted. High-energy proton spallation on $^{232}Th$ can provide recoverable $^{228}Th$ as an alternative source for generators.[97]

## Scandium-44

### Radioisotope Overview

$^{44}Sc$ emits $\beta^+$ particles with a 4·0 h half-life.[1] High-energy 1157 keV γ-rays are co-emitted.[1]

### Clinical Status

$^{44}Sc$ is in the early stages of clinical validation as a PET isotope.[98] Its half-life provides more time for transportation and imaging compared to $^{68}Ga$.[10] $^{44}Sc$'s emission of both $\beta^+$ and γ-rays makes it a candidate for 3γ imaging.[99]

Like $^{177}Lu$ and $^{90}Y$, $^{44}Sc$ can coordinate stably to the common chelator DOTA and has been successfully used as a PET analogue to these radioisotopes in human trials.[10] The chemically matched theranostic pair $^{44}Sc/^{47}Sc$ is promising.

### Production Mechanisms

$^{44}Sc$ can be produced with reasonable purity with a biomedical proton cyclotron irradiating a $^{44}Ca$ target or using a $^{44}Ti/^{44}Sc$ generator.[10] Irradiation of $^{44}Ca$ has been used to supply a clinical trial.[98] Use of the generator is limited by cyclotron-based production of $^{44}Ti$, the activity of which builds at a slow rate due to its very long half-life of 59·1 y,[1] and generator chemistry requires further development. High-current cyclotrons have potential to increase generator production.

## Scandium-47

### Radioisotope Overview

$^{47}Sc$ decays by $\beta^-$ emission with a half-life of 3·35 d and mean and maximum β energies of 0·162 MeV and 0·6 MeV. It co-emits 159 keV γ-rays.[1]

### Clinical Status

$^{47}Sc$ shows promise for $\beta^-$ therapy. Compared to $^{177}Lu$ it has lower energies and a shorter half-life. Its γ-rays are suitable for SPECT.[10] $^{47}Sc$ has two chemically matched diagnostic partners, $^{44}Sc$ and $^{43}Sc$. Preclinical studies have shown $^{47}Sc$'s potential,[10] but no clinical studies have been completed to date.

### Production Mechanisms

$^{47}Sc$ production using reactors, cyclotrons, and e-linacs has been explored. Co-production of contaminants $^{48}Sc$ and $^{46}Sc$ makes many production routes not ideal for radiopharmaceutical purposes. Producing $^{47}Ca$ that shortly decays to $^{47}Sc$ is one method that ensures radionuclidic purity of $^{47}Sc$. High yields of this pseudo-generator system have been reported from irradiating $^{46}Ca$ in a high-flux reactor and from the photonuclear reaction on $^{48}Ca$.[69] $^{47}Sc$ has recently been produced with manageable levels of contaminants and good yield by irradiating natural vanadium with protons.[100]

## Terbium-149

### Radioisotope Overview

$^{149}Tb$ decays by α emission (16·7%; 3·97 MeV), positron emission (7·1%; mean and maximum positron energies of 0·73 MeV and 2·45 MeV), and electron capture (76·2%), with a 4·12 h half-life. Unlike many α emitters, $^{149}Tb$ produces no α-emitting daughter cascade.[1]

### Clinical Status

$^{149}Tb$ remains preclinical but promising for TAT. Targeted therapeutic effects have been demonstrated using rituximab and folate-, PSMA-, and SSTR-targeted agents, while allowing simultaneous PET imaging. Limited availability has prevented clinical evaluation.[101–104]

### Production Mechanisms

Current production uses proton spallation on tantalum or $^{165}Ho$ followed by mass separation and chemical purification.[104] A lower-energy alternative could use a proton beam on $^{152}Gd$, although co-produced terbium isotopes still require mass separation.

## Terbium-152

### Radioisotope Overview

$^{152}Tb$ decays by positron emission (20·3%; mean and maximum positron energies of 1·14 MeV and 2·62 MeV) and electron capture (79·7%) to $^{152}Gd$, with a 17·5 h half-life.[1]

### Clinical Status

$^{152}$Tb remains investigational, but human PET imaging has been demonstrated with SSTR-targeted DOTATOC and PSMA-targeted PSMA-617.[102,105] It is the diagnostic member of the terbium family and can be used as a theranostic partner of $^{149}$Tb and $^{161}$Tb.

### Production Mechanisms

Current production uses proton spallation of tantalum followed by mass separation and chemical purification.[106] A lower-energy alternative uses protons on enriched $^{152}$Gd. This could enable production with conventional medical cyclotrons, although highly enriched $^{152}$Gd, currently not available at sufficiently high enrichment levels, is required to suppress $^{153}$Tb contamination.[107]

## Terbium-161

### Radioisotope Overview

$^{161}$Tb decays by $\beta^-$ emission to $^{161}$Dy with a 6·95-d half-life and mean and maximum β energies of 0·15 and 0·59 MeV. It also emits short-range conversion and Auger electrons alongside 49 and 75-keV photons.[1]

### Clinical Status

$^{161}$Tb remains investigational, but human use has been demonstrated for neuroendocrine tumours and prostate cancer.[105,108] Its additional short-range electron emissions may improve treatment of micro-metastases and individual tumour cells compared to $^{177}$Lu due to higher local dose density.

### Production Mechanisms

$^{161}$Tb is produced by neutron irradiation of enriched $^{160}$Gd producing $^{161}$Gd, followed by $\beta^-$ decay and chemical separation to yield no-carrier-added $^{161}$Tb.[109] We have identified no compelling alternative route.

## Thorium-227

### Radioisotope Overview

$^{227}$Th decays by α emission with an 18·7-day half-life and mean energy of 5·9 MeV. Its decay chain proceeds through $^{223}$Ra to $^{207}$Pb, emitting five α particles in total.[1]

### Clinical Status

$^{227}$Th remains investigational. Stable 3,2-HOPO chelators permit attachment to antibodies, and targeted $^{227}$Th conjugates are in early-phase human trials for CD22, mesothelin, PSMA, and HER2 targets.[110,111]

### Production Mechanisms

$^{227}$Th is separated from long-lived $^{227}$Ac generators. $^{227}$Ac is conventionally produced by reactor irradiation of $^{226}$Ra to produce $^{227}$Ra, followed by $\beta^-$ decay.[45] Accelerator-driven neutron systems could provide supplementary production capacity. Alternatively, natural thorium irradiated with

high-energy protons produces $^{227}Th$ directly and co-produces the $^{227}Ac$ generator. This can provide substantial yields but requires a centralised spallation facility.[112]

# Appendix B: Isotope Nuclear Production Pathways

## The Established Clinical Core

### Carbon-11

| Mechanism | Required Beam | Yield | Challenges | References |
|---|---|---|---|---|
| $^{14}N(p,\alpha)^{11}C$ | 13→3 MeV protons | >100 GBq (typical batch yield) | Requires high-purity target gas, sensitive to sources of carrier carbon | [113, 16] |

Table B1: Key nuclear reactions identified to generate Carbon-11

### Chromium-51

| Mechanism | Required Beam | Yield | Challenges | References |
|---|---|---|---|---|
| $^{50}Cr(n,\gamma)^{51}Cr$ | Thermal neutrons | σ=15·9 b | Required enriched target, carrier-added product | [6] |
| $^{nat}V(p,n)^{51}Cr$ | 1·6+ MeV protons (peak σ at 12 MeV) | 22 MBq/µAh (30 MeV) | May require cooling time for co-produced $^{49}Cr$ | [114, 115] |
| $^{nat}V(d,2n)^{51}Cr$ | 3·9+ MeV deuterons (peak σ at 17 MeV) | 26 MBq/µAh (30 MeV) | Requires specialised facility, may require cooling time for co-produced $^{49}Cr$ | [114, 116] |

Table B2: Key nuclear reactions identified to generate Chromium-51

### Cobalt-60

| Mechanism | Required Beam | Yield | Challenges | References |
|---|---|---|---|---|
| $^{nat}Co(n,\gamma)^{60}Co$ | Thermal neutrons | σ=37·18 b | Carrier-added product | [6] |

Table B3: Key nuclear reactions identified to generate Cobalt-60

## Copper-64

| Mechanism | Required Beam | Yield | Challenges | References |
|---|---|---|---|---|
| $^{64}Ni(p,n)^{64}Cu$ | 12→8 MeV protons | 304 MBq/µAh | Target recycling required | 117 |
| $^{64}Ni(d,2n)^{64}Cu$ | 17→11 MeV deuterons | 430 MBq/µAh | Specialised facility, target recycling required | 117 |
| $^{68}Zn(p,\alpha n)^{64}Cu$ | 30→21 MeV protons | 116 MBq/µAh | Dominant $^{67}Ga$ co-production, target recycling required | 117, 118 |
| $^{66}Zn(p,2pn)^{64}Cu$ | 52→37 MeV protons | 316 MBq/µAh | Specialised facility, $^{61}Cu$ co-production, $^{65}Zn$ accumulation in recycled target | 117, 118 |
| $^{64}Zn(d,2p)^{64}Cu$ | 20→10 MeV deuterons | 27·1 MBq/µAh | Specialised facility, low production rate, $^{61}Cu$ co-production, $^{65}Zn$ accumulation in recycled target | 117, 118 |
| $^{66}Zn(d,\alpha)^{64}Cu$ | 13→5 MeV deuterons | 13·8 MBq/µAh | Specialised facility, low production rate, $^{65}Zn$ accumulation in recycled target | 117,118 |
| $^{nat}Zn(d,x)^{64}Cu$ | 25→10 MeV deuterons | 57·0 MBq/µAh | Specialised facility, $^{61}Cu$ co-production | 117,118 |

Table B4: Key nuclear reactions identified to generate Copper-64

## Fluorine-18

| Mechanism | Required Beam | Yield | Challenges | References |
|---|---|---|---|---|
| $^{18}O(p,n)^{18}F$ | 16→3 MeV protons | 150 GBq (typical batch yield) | Requires enriched water target, co-production of 13N at lower enrichment, current limited by liquid target vaporisation and cavitation | 16,113 |
| $^{nat}Ne(d,x)^{18}F$ | 15→1·5 MeV deuterons | 3·61 GBq/µA (saturation yield) | Specialised facility | 16 |

Table B5: Key nuclear reactions identified to generate Fluorine-18

## Gallium-67

| Mechanism | Required Beam | Yield | Challenges | References |
|---|---|---|---|---|
| $^{68}Zn(p,2n)^{67}Ga$ | 26→18 MeV protons | 50 GBq (typical batch yield) | Requires enriched target, co-production of $^{66}Ga$, requires cooling time | 16,113 |
| $^{67}Zn(p,n)^{67}Ga$ | 15→8 MeV protons | 3 GBq (typical batch yield) | Requires highly enriched target, lower yield, co-production of $^{66}Ga$ | 58,113 |
| $^{66}Zn(d,n)^{67}Ga$ | 10→2 MeV deuterons | 11·63 MBq/μAh | Specialised facility, co-production of $^{66}Ga$, requires target recovery | 119,120 |

Table B6: Key nuclear reactions identified to generate Gallium-67

## Gallium-68

| Mechanism | Required Beam | Yield | Challenges | References |
|---|---|---|---|---|
| $^{nat}Ga(p,xn)^{68}Ge \rightarrow ^{68}Ga$ | 70→20 MeV protons | 40 GBq $^{68}Ge$ (typical batch yield) | Specialised facility, low production rate of parent isotope, co-production of $^{71}Ge$, significant co-production of non-isotopic radioisotopes, low yield rate from generator | 10,113,121 |
| $^{nat}Ga(p,xn)^{68}Ge \rightarrow ^{68}Ga$ | 30→10 MeV protons | 1·5 MBq/μAh $^{68}Ge$ | Low production rate of parent isotope, co-production of $^{71}Ge$, significant co-production of non-isotopic radioisotopes, low yield rate from generator | 10,58,121 |
| $^{69}Ga(p,2n)^{68}Ge \rightarrow ^{68}Ga$ | 10+ MeV protons | 2·3 MBq/μAh $^{68}Ge$ | Requires enriched target | 122 |
| $^{68}Zn(p,n)^{68}Ga$ | <12 MeV protons | 599 MBq/μAh (14·2 MeV) | Requires enriched target | 58,113,123 |
| $^{65}Cu(\alpha,n)^{68}Ga$ | 25→5 MeV α particles | 500 MBq/μAh | Specialised facility, requires enriched target | 113,122 |

Table B7: Key nuclear reactions identified to generate Gallium-68

## Indium-111

| Mechanism | Required Beam | Yield | Challenges | References |
|---|---|---|---|---|
| $^{112}Cd(p,2n)^{111}In$ | 25→18 MeV protons | 50 GBq (typical batch yield) | Requires target recycling, co-produced $^{114m}In$ | [16,113] |
| $^{111}Cd(p,n)^{111}In$ | 25→18 MeV protons | 3 GBq (typical batch yield) | Requires target recycling, co-produced $^{114m}In$ | [16,113] |
| $^{109}Ag(\alpha,2n)^{111}In$ | 35→17 MeV α particles | 1000 MBq/μAh | Specialised facility, expensive enriched target | [16,124] |

Table B8: Key nuclear reactions identified to generate Indium-111

## Iodine-123

| Mechanism | Required Beam | Yield | Challenges | References |
|---|---|---|---|---|
| $^{124}Xe(p,2n)^{123}Cs \rightarrow {}^{123}Xe \rightarrow {}^{123}I$ | 29→23 MeV protons | 70 GBq typical batch | Requires target recycling, $^{121}I(\rightarrow {}^{121}Te)$ co-production | [113,125] |
| $^{123}Te(p,n)^{123}I$ | 14·5→10 MeV protons | 20 GBq typical batch | Requires target recycling | [113] |
| $^{127}I(p,5n)^{123}Xe \rightarrow {}^{123}I$ | 65→45 MeV protons | 40 GBq typical batch | $^{125}Xe(\rightarrow {}^{125}I)$ co-production | [113] |
| $^{124}Xe(\gamma,n)^{123}Xe \rightarrow {}^{123}I$ | Photon source | 0·14 MBq/(g μAh) (40 MeV e) | Specialised facility, requires target recycling | [126] |

Table B9: Key nuclear reactions identified to generate Iodine-123

## Iodine-124

| Mechanism | Required Beam | Yield | Challenges | References |
|---|---|---|---|---|
| $^{124}Te(p,n)^{124}I$ | 12→8 MeV protons | 16 MBq/μAh | $^{125}I$ co-production | [113] |
| $^{125}Te(p,2n)^{124}I$ | 22→15 MeV protons | 93 MBq/μAh | Requires cooling time | [113] |
| $^{124}Te(d,2n)^{124}I$ | 14→10 MeV deuterons | 17·5 MBq/μAh | $^{125}I$ co-production | [127] |
| $^{123}Sb(\alpha,3n)^{124}I$ | 45→32 MeV alpha particles | 15·5 MBq/μAh | Specialised facility, requires cooling time, $^{125}I$ co-production | [128] |

Table B10: Key nuclear reactions identified to generate Iodine-124

## Iodine-125

| Mechanism | Required Beam | Yield | Challenges | References |
|---|---|---|---|---|
| $^{124}Xe(n,\gamma)^{125}Xe \rightarrow {}^{125}I$ | Thermal neutrons | σ=114 b, I=2300 b | Requires target recycling | 6,129 |
| $^{nat}Xe(n,x)^{125}Xe \rightarrow {}^{125}I$ | Thermal neutrons | σ=114 b, I=2300 b ($^{124}Xe$, 0·095% $^{nat}Xe$) | Requires long irradiation time, co-produced $^{126}I$ | 6 |

Table B11: Key nuclear reactions identified to generate Iodine-125

## Iodine-131

| Mechanism | Required Beam | Yield | Challenges | References |
|---|---|---|---|---|
| $^{235}U(n,f)^{131}I$, $^{235}U(n,f)\ ^{131}Te \rightarrow {}^{131}I$ | Thermal neutrons / 14 MeV neutrons | 2·88% fission yield (thermal) / 4·09% fission yield (14 MeV) | Expensive processing, requires optimisation of cooling time, reliance on HEU, radioactive waste | 129 |
| $^{130}Te(n,\gamma)^{131}Te \rightarrow {}^{131}I$ | Thermal neutrons | σ=0·2 b, I=0·239 b | Low cross-section | 129 |

Table B12: Key nuclear reactions identified to generate Iodine-131

## Iridium-192

| Mechanism | Required Beam | Yield | Challenges | References |
|---|---|---|---|---|
| $^{191}Ir(n,\gamma)^{192}Ir$ | Thermal neutrons | σ=962 b, I=1132 b | Carrier-added product | 129 |
| $^{192}Os(p,n)^{192}Ir$ | 16→8 MeV protons | 0·16 MBq/μAh | Low yield, target recovery required, $^{192m2}Ir$ co-production | 130 |
| $^{192}Os(d,2n)^{192}Ir$ | 4·1+ MeV deuterons (peak σ at 12 MeV) | 0·9 MBq/μAh (20 MeV) | $^{192m2}Ir$ co-production (higher than (p,n) route) | 131 |

Table B13: Key nuclear reactions identified to generate Iridium-192

## Krypton-81m

| Mechanism | Required Beam | Yield | Challenges | References |
|---|---|---|---|---|
| $^{nat}Kr(p,x)^{81}Rb \rightarrow {}^{81m}Kr$ | 28→20 MeV protons | 1946 MBq/μAh | Low thermal conductivity of $^{nat}Kr$ gas | 16,132,133 |

Table B14: Key nuclear reactions identified to generate Krypton-81m

## Lutetium-177

| Mechanism | Required Beam | Yield | Challenges | References |
|---|---|---|---|---|
| $^{176}Yb(n,\gamma)^{177}Yb \rightarrow {}^{177}Lu$ | Thermal neutrons | σ=2·85 b, I=6·77 b | Low capture cross section, requires target recycling, complex separation chemistry | 129,134 |
| $^{176}Lu(n,\gamma)^{177}Lu$ | Thermal neutrons | σ=2097 b, I=9190·9 b | Carrier-added production, $^{177m}Lu$ co-production | 129,134 |
| $^{176}Yb(d,p)^{177}Yb \rightarrow {}^{177}Lu$ | <20 MeV deuterons (peak σ at 12·5 MeV) | 0·83 MBq/μA·h (20 MeV) | Specialised facility, requires target recycling, complex separation chemistry | 129,135,136 |
| $^{179}Hf(\gamma,pn)^{177}Lu$ | Photon source | 0·19 MBq/μAh (45 MeV e) | Specialised facility, $^{178}Lu$ co-production | 137 |
| $^{178}Hf(\gamma,p)^{177}Lu$ | Photon source | 0·08 MBq/μAh (45 MeV e) | Specialised facility, lower yields | 137 |

Table B15: Key nuclear reactions identified to generate Lutetium-177

## Nitrogen-13

| Mechanism | Required Beam | Yield | Challenges | References |
|---|---|---|---|---|
| $^{16}O(p,\alpha)^{13}N$ | 16→7 MeV protons | 30 GBq (typical batch yield) | Requires sophisticated heat removal and steam condensation systems, pyrogen growth in targets | 16,113,138 |
| $^{13}C(p,n)^{13}N$ | 15→8 MeV protons | 10 GBq (typical batch yield) | Requires highly enriched target | 113 |
| $^{nat}C(d,x)^{13}N$ | 6→1 MeV deuterons | - | Requires specialised facility | 16 |
| $^{14}N(d,x)^{13}N$ | 4·9+ MeV deuterons (peak σ at 10 MeV) | 90 MBq/μA (saturation yield, 12 MeV) | Production is a byproduct of $^{15}O$ production | 113,139 |

Table B16: Key nuclear reactions identified to generate Nitrogen-13

## Oxygen-15

| Mechanism | Required Beam | Yield | Challenges | References |
|---|---|---|---|---|
| $^{14}N(d,n)^{15}O$ | 8→0 MeV deuterons | 80 GBq (typical batch yield) | Specialised facility | [113] |
| $^{15}N(p,n)^{15}O$ | 12→0 MeV protons | 80 GBq (typical batch yield) | Requires highly enriched target | [113] |
| $^{nat}O(\gamma,n)^{15}O$ | Photon source | - | Boiling of water target | [41] |
| $^{nat}O(p,pn)^{15}O$ | 26·5→20 MeV protons | - | Cavitation of liquid target | [16,140] |

Table B17: Key nuclear reactions identified to generate Oxygen-15

## Palladium-103

| Mechanism | Required Beam | Yield | Challenges | References |
|---|---|---|---|---|
| $^{102}Pd(n,\gamma)^{103}Pd$ | Thermal neutrons | σ=1·88 b, I=23·0 b | Carrier-added product | [129] |
| $^{103}Rh(p,n)^{103}Pd$ | 13→7 MeV protons | 50 GBq (typical batch yield) | Difficult separation chemistry, expensive target | [113,141] |
| $^{103}Rh(d,2n)^{103}Pd$ | 13·3→3·6 MeV deuterons | 9·4 MBq/μAh | Specialised facility, difficult separation chemistry, expensive target | [142] |
| $^{107}Ag(p,x)^{103}Pd$ | 80→30 MeV protons | - | Specialised facility, $^{100}Pd$ co-production | [113,143] |

Table B18: Key nuclear reactions identified to generate Palladium-103

## Radium-223

| Mechanism | Required Beam | Yield | Challenges | References |
|---|---|---|---|---|
| $^{nat}U \rightarrow ^{223}Ra$ / $^{231}Pa \rightarrow ^{223}Ra$ | None (decay of uranium ore) | - | Limited source supply | 144 |
| $^{226}Ra(n,\gamma)^{227}Ra \rightarrow ^{227}Ac \rightarrow ^{223}Ra$ | Thermal neutrons | σ=13 b | Limited target supply, target decays to $^{222}Rn$, minimise fast neutrons to avoid target fission | 144 |
| $^{nat}Th(p,x)^{227}Th \rightarrow ^{223}Ra$, $^{nat}Th(p,x)^{223}Ra$ | 800 MeV | 1·6 MBq/µAh $^{227}Th$, 1·1 MBq/µAh $^{223}Ra$ | Specialised facility, low yield, radioactive waste, complex chemistry | 145 |
| $^{227}Ac \rightarrow ^{223}Ra$ | None (decay of legacy $^{227}Ac$ sources) | - | Limited source supply, requires sophisticated purification chemistry | 144,146 |

Table B19: Key nuclear reactions identified to generate Radium-223

## Rubidium-82

| Mechanism | Required Beam | Yield | Challenges | References |
|---|---|---|---|---|
| $^{nat}Rb(p,xn)^{82}Sr \rightarrow ^{82}Rb$ | 70→50 MeV protons | 70 GBq $^{82}Sr$ (typical batch yield) | Specialised facility, co-produced $^{85}Sr$ | 113 |
| $^{nat}Mo(p,x)^{82}Sr \rightarrow ^{82}Rb$ | 500+ MeV protons | 3·7 MBq/µAh (800 MeV) | Specialised facility | 121 |
| $^{nat}Kr(\alpha,2n)^{82}Sr \rightarrow ^{82}Rb$ | 120→20 MeV alpha particles | 0·96 MBq/µAh | Specialised facility, co-produced $^{83,85}Sr$ (more than ($^{3}He$,x) route), low yield rate | 121 |
| $^{nat}Kr(^{3}He,x)^{82}Sr$ | 90→20 MeV $^{3}He$ | 1·3 MBq/µAh | Specialised facility (rarer than α facility), co-produced $^{83,85}Sr$ requires He gas recovery, low yield rate, high $^{3}He$ currents not available | 121 |

Table B20: Key nuclear reactions identified to generate Rubidium-82

## Selenium-75

| Mechanism | Required Beam | Yield | Challenges | References |
|---|---|---|---|---|
| $^{74}Se(n,\gamma)^{75}Se$ | Thermal and epithermal neutrons | σ=48·2 b, I=515 b | Carrier-added product | [147] |
| $^{75}As(p,n)^{75}Se$ | 18→8 MeV protons | 2 MBq/μAh | Low yield rate | [148] |
| $^{75}As(d,2n)^{75}Se$ | <30 MeV deuterons | 3·5 MBq/μAh (30 MeV) | Specialised facility, low yield rate | [148] |

Table B21: Key nuclear reactions identified to generate Selenium-75

## Strontium-89

| Mechanism | Required Beam | Yield | Challenges | References |
|---|---|---|---|---|
| $^{89}Y(n,p)^{89}Sr$ | Fast neutrons | 20 GBq (typical batch yield, high-flux reactor) | Requires high fast neutron flux | [113] |
| $^{88}Sr(n,\gamma)^{89}Sr$ | Thermal neutrons | σ=5·9 × $10^{-5}$ b | Requires very high flux reactor, requires enriched target to minimise co-produced $^{85}Sr$, carrier-added product | [149] |

Table B22: Key nuclear reactions identified to generate Strontium-89

## Technetium-99m

| Mechanism | Required Beam | Yield | Challenges | References |
|---|---|---|---|---|
| $^{235}U(n,f)^{99}Mo \rightarrow ^{99m}Tc$ | Thermal neutrons | >$10^3$ GBq (typical batch yield) | Reliance on HEU, radioactive waste, expensive processing | 113 |
| $^{100}Mo(p,2n)^{99m}Tc$ | 22→10 MeV protons | 630 MBq/μAh | Requires target recycling, co-produced $^{97m,g,98,99g}Tc$ | 58,113,114 |
| $^{98}Mo(n,\gamma)^{99}Mo \rightarrow ^{99m}Tc$ | Thermal neutrons | 500 GBq (typical batch yield in high-flux reactor) | Requires target recycling | 113,150 |
| $^{100}Mo(d,x)^{99}Mo \rightarrow ^{99m}Tc$ | 2·1+ MeV deuterons | 1900 MBq/ μAh $^{99}Mo$ (45 MeV) | Specialised facility | 114 |
| $^{100}Mo(d,3n)^{99m}Tc$ | 10+ MeV deuterons (peak σ at 22 MeV) | 750 MBq/ μAh | Specialised facility, lower yield, requires energy optimisation to avoid co-produced $^{98}Tc$ | 114 |
| $^{238}U(\gamma,f)^{99}Mo$ | Photon source (peak σ at 14 MeV) | - | Specialised facility, reliance on HEU, radioactive waste, expensive processing | 114 |

Table B23: Key nuclear reactions identified to generate Technetium-99m (Molybdenum-99 parent)

## Thallium-201

| Mechanism | Required Beam | Yield | Challenges | References |
|---|---|---|---|---|
| $^{203}Tl(p,3n)^{201}Pb \rightarrow ^{201}Tl$ | 28→20 MeV protons | 70 GBq (typical batch yield, 32 h EOB) | Requires high target enrichment, requires cooling time, requires two separation steps, co-produced $^{200}Pb(\rightarrow ^{200}Tl)$, $^{202m}Pb(\rightarrow ^{202}Tl)$ | 58,113,114 |
| $^{nat}Tl(p,xn)^{201}Pb \rightarrow ^{201}Tl$ | 28→20 MeV protons | 30 GBq (typical batch yield, 32 h EOB) | Requires cooling time, requires two separation steps, co-produced $^{200}Pb(\rightarrow ^{200}Tl)$, $^{202m}Pb(\rightarrow ^{202}Tl)$ | 58,113,114 |

Table B25: Key nuclear reactions identified to generate Thallium-201

## Xenon-133

| Mechanism | Required Beam | Yield | Challenges | References |
|---|---|---|---|---|
| $^{235}U(n,f)^{133}Xe$ | Thermal neutrons | 6·7% fission yield | Reliance on HEU, radioactive waste, expensive processing | 6 |

Table B26: Key nuclear reactions identified to generate Xenon-133

## Yttrium-90

| Mechanism | Required Beam | Yield | Challenges | References |
|---|---|---|---|---|
| $^{235}U(n,f)^{90}Sr \rightarrow ^{90}Y$ | Thermal neutrons | 5·7% thermal fission yield | Reliance on HEU, radioactive waste, expensive processing | 6 |
| $^{89}Y(n,\gamma)^{90}Y$ | Thermal neutrons | σ=1·28 b, I=0·46 b | Carrier-added product | 129 |

Table B27: Key nuclear reactions identified to generate Yttrium-90

## Zirconium-89

| Mechanism | Required Beam | Yield | Challenges | References |
|---|---|---|---|---|
| $^{89}Y(p,n)^{89}Zr$ | 14→9 MeV protons | 58 MBq/µAh | Solid target (non-routine for clinics) | 113 |

Table B28: Key nuclear reactions identified to generate Zirconium-89

# The Underleveraged Middle

## Erbium-169

| Mechanism | Required Beam | Yield | Challenges | References |
|---|---|---|---|---|
| $^{168}Er(n,\gamma)^{169}Er$ | Thermal neutrons | σ=2·3 b | Low capture cross section, carrier-added product | 70 |
| $^{169}Tm(n,p)^{169}Er$ | Spallation neutrons | - | Specialised facility, expensive target | 113 |

Table B29: Key nuclear reactions identified to generate Erbium-169

## Gold-198

| Mechanism | Required Beam | Yield | Challenges | References |
|---|---|---|---|---|
| $^{197}Au(n,\gamma)^{198}Au$ | Thermal and epithermal neutrons | σ=98·65 b | Carrier-added product | 6,71 |
| $^{198}Pt(d,2n)^{198}Au$ | 18→7 MeV deuterons | 16·0 MBq/µAh (15 MeV) | Specialised facility, expensive target, $^{199}Au$ co-production | 71,151 |
| $^{198}$Pt(p,n)$^{198}$Au | 16→6 MeV protons | 4·3 MBq/μAh | Expensive target, $^{196}$Au co-production | 71,151 |
| $^{nat}$Pt(d,x)$^{198}$Au | 5+ MeV deuterons (peak σ at 13 MeV) | 1·7 MBq/μAh (15 MeV) | Specialised facility, $^{196,199}$Au co-production | 152 |
| $^{nat}$Hg(γ,p)$^{198}$Au | Photon source | - | Specialised facility, mercury toxicity, target volatility, difficult separation chemistry | 71 |

Table B30: Key nuclear reactions identified to generate Gold-198

## Rhenium-186

| Mechanism | Required Beam | Yield | Challenges | References |
|---|---|---|---|---|
| $^{185}$Re(n,γ)$^{186}$Re | Thermal neutrons | σ=112 b, I=1729 b | Carrier-added product | 129 |
| $^{186}$W(p,n)$^{186}$Re | <18 MeV protons | 4 MBq/μAh (18 MeV) | Requires target recycling, co-produced $^{184m,g}$Re and stable Re, low yield | 113,153 |
| $^{186}$W(d,2n)$^{186}$Re | <20 MeV deuterons | 14 MBq/μAh (18 MeV) | Requires target recycling, co-produced $^{184m,g}$Re and stable Re | 113,153 |

Table B31: Key nuclear reactions identified to generate Rhenium-186

## Rhenium-188

| Mechanism | Required Beam | Yield | Challenges | References |
|---|---|---|---|---|
| $^{186}$W(n,γ)$^{187}$W(n,γ) $^{188}$W→$^{188}$Re | Thermal neutrons | 20 GBq (typical batch yield in a high-flux reactor) | Low cross-sections, second capture competes with $^{187}$W decay, few suitable reactors | 113 |
| $^{nat}$Os(γ,x)$^{188}$Re | Photon source | - | Specialised facility, co-produced $^{189}$Re (may not pose a problem; decay properties are similar to $^{188}$Re), co-produced $^{184}$Re (<1% activity) | 154 |

Table B32: Key nuclear reactions identified to generate Rhenium-188

## Samarium-153

| Mechanism | Required Beam | Yield | Challenges | References |
|---|---|---|---|---|
| $^{152}Sm(n,\gamma)^{153}Sm$ | Thermal neutrons | σ=213 b, I= 2940 b | Requires enriched target, carrier-added product | 129 |
| $^{nat}Sm(n,x)^{153}Sm$ | Thermal neutrons | σ=213 b, I= 2940 b | Carrier added product, lower specific activity | 129 |
| $^{150}Nd(\alpha,n)^{153}Sm$ | 25→15 MeV alpha particles | 1·1 MBq/μAh | Specialised facility, high target enrichment required | 155 |

Table B33: Key nuclear reactions identified to generate Samarium-153

# The Emerging Frontier

## Actinium-225

| Mechanism | Required Beam | Yield | Challenges | References |
|---|---|---|---|---|
| $^{229}Th \rightarrow {}^{225}Ac$ | None (generator stocks) | 63 GBq/year (worldwide total) | Inadequate $^{229}Th$ supply | 83 |
| $^{226}Ra(\gamma,n)^{225}Ra \rightarrow {}^{225}Ac$ | Photon source >6·4 MeV | 335·4 MBq/μAh/(g $^{226}Ra$) (39 MeV, 105 mA) | Specialised facility, radioactive target, target produces $^{222}Rn$ gas, co-produced $^{227}Ac$ (first $^{226}Ra/^{225}Ra$ elution must be discarded) | 83,84 |
| $^{226}Ra(n,2n)^{225}Ra \rightarrow {}^{225}Ac$ | Fast neutrons | 15·4 GBq/(g $^{226}Ra$) $^{225}Ac$ (45 d irradiation in fast reactor), 370-1110 MBq/(g $^{226}Ra$) $^{225}Ac$ per week (secondary neutrons) | Specialised facility, radioactive target, target produces $^{222}Rn$ gas, co-produced $^{227}Ra(\rightarrow {}^{227}Ac)$ when produced in a reactor | 156,157 |
| $^{232}Th(p,x)^{225}Ac$, $^{232}Th(p,x)^{225}Ra \rightarrow {}^{225}Ac$ | 140→60 MeV protons | 4 MBq/μAh | Specialised facility, co-produced $^{227}Ac$, complex chemical processing, radioactive waste | 83,113 |
| $^{226}Ra(p,2n)^{225}Ac$ | 22→10 MeV protons | 7 MBq/μAh | Specialised facility, radioactive target, target produces $^{222}Rn$ gas, complicated chemical processing, co-produced $^{226}Ac$ and possible $^{227}Ac$ | 83,113 |

Table B34: Key nuclear reactions identified to generate Actinium-225

## Astatine-211

| Mechanism | Required Beam | Yield | Challenges | References |
|---|---|---|---|---|
| $^{209}Bi(\alpha,2n)^{211}At$ | 29→20·7 MeV alpha particles | 16·3-41 MBq/µAh reported (28-29·5 MeV incident alpha) | Specialised facility, energy capped to avoid $^{210}At(\rightarrow ^{210}Po)$ co-production, low thermal conductivity of target, low melting point of target | 90,129 |
| $^{209}Bi(^{7}Li, 5n)^{211}Rn \rightarrow ^{211}At$ | 40+ MeV $^{7}Li$ (peak σ at 54 MeV) | 29 MBq $^{211}At$ (50 MeV, 1 pµA, 10 h) | Specialised facility, $^{210}Rn(\rightarrow ^{210}At)$ co-production | 158 |

Table B35: Key nuclear reactions identified to generate Astatine-211

## Bismuth-213

| Mechanism | Required Beam | Yield | Challenges | References |
|---|---|---|---|---|
| $^{225}Ac \rightarrow ^{213}Bi$ | None | - | Inadequate $^{225}Ac$ supply | 10 |

Table B36: Key nuclear reactions identified to generate Bismuth-213

## Copper-67

| Mechanism | Required Beam | Yield | Challenges | References |
|---|---|---|---|---|
| $^{68}Zn(\gamma,p)^{67}Cu$ | Photon source | 0·2 MBq/(µA·h·g/cm²) (55 MeV e, 0·3 µA) | Specialised facility, target recycling required | 159 |
| $^{70}Zn(d,x)^{67}Cu$ | 26→16 MeV deuterons | 6·4 MBq/µAh | Specialised facility, low production rate, target recycling required | 92 |
| $^{68}Zn(p,2p)^{67}Cu$ | 80→30 MeV protons | 42 MBq/µAh | Specialised facility, $^{64}Cu$ co-production, target recycling required | 113,160,161 |
| $^{70}Zn(p,\alpha)^{67}Cu$ | 18→12 MeV protons | 2·2 MBq/µAh | Low production rate, $^{64}Cu$ co-production, target recycling required | 113,160 |
| $^{67}Zn(n,p)^{67}Cu$ | Fission spectrum | 4·8 × 10³ MBq/(g·h) | $^{64}Cu$ co-production, target recycling required | 113 |

Table B37: Key nuclear reactions identified to generate Copper-67

## Lead-212

| Mechanism | Required Beam | Yield | Challenges | References |
|---|---|---|---|---|
| $^{232}Th \rightarrow ^{228}Th \rightarrow ^{224}Ra \rightarrow ^{212}Pb$ | None ($^{232}Th$ stockpiles) | 3·7 GBq $^{228}Ra$ per ton of aged $^{232}Th$ | Large volumes of material require chemical processing | 45 |
| $^{232}U \rightarrow ^{228}Th \rightarrow ^{224}Ra \rightarrow ^{212}Pb$ | None ($^{232}U$ stockpiles) | - | Limited supply of $^{232}U$, complex chemical processing | 45 |
| $^{226}Ra(n,3n)^{224}Ra \rightarrow ^{212}Pb$ | Fast neutrons | 185-2220 MBq /(g $^{226}Ra$) $^{228}Ra$ per week (secondary neutrons) | Specialised facility, radioactive target, target produces $^{222}Rn$ gas | 157 |

Table B38: Key nuclear reactions identified to generate Lead-212

## Scandium-44

| Mechanism | Required Beam | Yield | Challenges | References |
|---|---|---|---|---|
| $^{44}Ca(p,n)^{44}Sc$ | 18→6 MeV protons | 2·3 × 10³ MBq/µAh | Target recycling required, difficult target handling | 113 |
| $^{44}Ca(d,2n)^{44}Sc$ | 25→10 MeV deuterons | 2·5 × 10³ MBq/µAh | Specialised facility, target recycling required, difficult target handling | 113 |
| $^{45}Sc(p,2n)^{44}Ti \rightarrow ^{44}Sc$ | 35→15 MeV protons | 0·004 MBq/µAh $^{44}Ti$ | Specialised facility, low production rate of generator, $^{44}Ti$ breakthrough during elution of $^{44}Sc$ from generator | 113,162 |
| $^{47}Ti(p,\alpha)^{44}Sc$ | 15→8 MeV protons | 48 MBq/µAh | Lower yield, $^{44m}Sc$ co-production | 163 |

Table B39: Key nuclear reactions identified to generate Scandium-44

## Scandium-47

| Mechanism | Required Beam | Yield | Challenges | References |
|---|---|---|---|---|
| $^{46}Ca(n,g)^{47}Ca \rightarrow ^{47}Sc$ | Thermal neutrons | σ=0·7 b | Low capture cross section, target recycling required | 164 |
| $^{48}Ca(\gamma,n)^{47}Ca \rightarrow ^{47}Sc$ | Photon source | 26 MBq/g (40 MeV e, 1mA, 6 d) | Specialised facility, target recycling required | 165 |
| $^{nat}V(p,x)^{47}Sc$ | 30→19 MeV protons | 82 MBq/µA | $^{46}Sc$ co-production | 166,167 |
| $^{nat}V(\gamma,x)^{47}Sc$ | Photon source | 4 MBq/g/h (38 MeV e, 0·2 mA) | Specialised facility, requires cooling time, $^{46}Sc$ co-production | 168 |
| $^{48}Ti(\gamma,p)^{47}Sc$ | Photon source <22·1 MeV | 0·26 MBq/g | Specialised facility, target recycling required | 169 |
| | | (20 MeV e, 1μA, 10 d) | | |
| $^{nat}Ca(p,2n)^{47}Sc$ | 24→17 MeV protons | 0·144 MBq/μAh | Low yield, $^{48}Sc$ co-production | 170 |
| $^{47}Ti(n,p)^{47}Sc$ | Fast neutrons | 7·8 MBq/mg ($10^{13}$ n $cm^{-2}$ $s^{-1}$ spallation neutrons) | Low yield, $^{46}Sc$ co-production | 164 |
| $^{50}Ti(p,\alpha)^{47}Sc$ | 22→10 MeV protons | 1·35 MBq/μAh (18 MeV, 0·28 mm target thickness) | Low yield, $^{46}Sc$ co-production (minimised through high target enrichment) | 171 |

Table B40: Key nuclear reactions identified to generate Scandium-47

## Terbium-149

| Mechanism | Required Beam | Yield | Challenges | References |
|---|---|---|---|---|
| $^{165}Ho(p,x)^{149}Tb$ | 1000 MeV protons | 5·8 MBq/μAh | Specialised facility, requires on-line mass separation, radioactive waste, complex chemical processing | 113 |
| $^{nat}Ta(p,x)^{149}Tb$ | 1000 MeV protons | 19 GBq/μA (100 $g/cm^2$ target) | Specialised facility, requires on-line mass separation, radioactive waste, complex chemical processing | 172 |
| $^{142}Nd(^{12}C,5n)^{149}Dy \rightarrow ^{149}Tb$ | 100→80 MeV $^{12}C$ | 2·5 MBq/μAh | Specialised facility, requires enriched target | 113 |
| $^{152}Gd(p,4n)^{149}Tb$ | 28+ MeV protons (peak σ at 41 MeV) | 2556 MBq/μAh (66 MeV) | Specialised facility, inadequate availability of enriched target, partial enrichment leads to significant co-produced Tb contaminants, difficult chemical separation from Gd | 173,174 |
| $^{152}Gd(\alpha,7n)^{149}Dy \rightarrow ^{149}Tb$ | 63+ MeV αparticles | 400 MBq/pμA (saturated yield, 100 MeV) | Specialised facility, high energy α beams generally unavailable, inadequate availability of enriched target, difficult chemical separation from Gd | 172,174 |

Table B41: Key nuclear reactions identified to generate Terbium-149

## Terbium-152

| Mechanism | Required Beam | Yield | Challenges | References |
|---|---|---|---|---|
| $^{nat}Ta(p,x)^{152}Tb$ | 1000 MeV protons | 77 MBq/µAh (1000 MeV) | Specialised facility, requires on-line mass separation, radioactive waste, complex chemical processing | 113 |
| $^{155}Gd(p,4n)^{152}Tb$ | 70→30 MeV protons | - | Specialised facility, requires highly enriched target | 113 |
| $^{nat}Nd(^{12}C,xn)^{152}Dy \rightarrow ^{152}Tb$ | 110→80 MeV $^{12}C$ | 10 MBq/µAh | Specialised facility | 113 |

Table B42: Key nuclear reactions identified to generate Terbium-152

## Terbium-161

| Mechanism | Required Beam | Yield | Challenges | References |
|---|---|---|---|---|
| $^{160}Gd(n,\gamma)^{161}Gd \rightarrow ^{161}Tb$ | Thermal neutrons | σ=1·4 b | Low capture cross section, difficult separation from decay product Dy | 90 |

Table B24: Key nuclear reactions identified to generate Terbium-161

## Thorium-227

| Mechanism | Required Beam | Yield | Challenges | References |
|---|---|---|---|---|
| $^{226}Ra(n,\gamma)^{227}Ra \rightarrow ^{227}Ac \rightarrow ^{227}Th$ | Thermal neutrons | σ=13 b | Limited target supply, target decays to radioactive $^{222}Rn$, fast neutrons should be minimised to avoid target fission | 144 |
| $^{nat}Th(p,x)^{227}Th$ | 800 MeV | 1·6 MBq/µAh | Specialised facility, low yield rates, radioactive waste, complex separation chemistry | 145 |
| $^{227}Ac \rightarrow ^{227}Th$ | None (decay of legacy $^{227}Ac$ sources) | - | Limited source supply, requires sophisticated purification chemistry | 146 |

Table B43: Key nuclear reactions identified to generate Thorium-227

# Appendix C: Additional Technology Intervention Opportunity Tables

| $^{A}X$ | Production constraint | Tech. Opportunity | Expected Outcome | Assessment |
|---|---|---|---|---|
| | | | | ***Established Clinical Core*** |
| $^{11}C$ | Well-managed | None | None | No technological intervention identified. |
| $^{51}Cr$ | Well-managed | None | None | No technological intervention identified. |
| $^{60}Co$ | Well-managed | None | None | No technological intervention identified. |
| $^{64}Cu$ | Well-managed | Investigate | Sustain | Produced with commercially available biomedical cyclotrons. Higher-power proton cyclotrons, using beam splitting across multiple characterised targets or improved target cooling, could increase aggregate production via $^{64}Ni(p,n)^{64}Cu$. |
| $^{18}F$ | Well-managed | None | None | Produced with commercially available biomedical cyclotrons. |
| $^{67}Ga$ | At-risk | None | None | No technological intervention identified. |
| $^{68}Ga$ | At-risk | Established | Expand | Produced with commercially available biomedical cyclotrons. Direct regional cyclotron production via $^{68}Zn(p,n)^{68}Ga$ can supplement generator-derived supply. |
| $^{111}In$ | Well-managed | None | None | No technological intervention identified. |
| $^{123}I$ | At-risk | None | None | No technological intervention identified. |
| $^{124}I$ | Sig. constrained | None | None | Produced with commercially available biomedical cyclotrons. Limited availability and PET reconstruction/quantification challenges have restricted clinical use. |
| $^{125}I$ | At-risk | Investigate | Sustain | Investigate production $^{124}Xe(n,\gamma)^{125}Xe \rightarrow ^{125}I$ from accelerator-driven thermal neutrons to augment reactor supply. |
| $^{131}I$ | At-risk | Investigate | Sustain | Produced routinely in reactors. Accelerator-driven thermal neutrons could augment reactor supply through $^{235}U(n,f)$, followed by recovery of fission-product $^{131}I$. |
| $^{192}Ir$ | Well-managed | None | None | Produced routinely in reactors. Production of higher specific activity forms is not well-motivated because a long half-life makes $^{192}Ir$ poorly suited for most targeted radiopharmaceuticals. |
| $^{81m}Kr$ | Sig. constrained | None | None | No technological intervention identified. Short generator half-life restricts distribution. |
| $^{177}Lu$ | At-risk | Investigate | Expand | Produced routinely in reactors. Deuteron irradiation of enriched ytterbium via $^{176}Yb(d,x)^{177}Lu$ merits investigation as an accelerator-based supplement to reactor production. Global demand, primarily from Lutathera and Pluvicto, is projected to continue increasing. |
| $^{13}N$ | Well-managed | None | None | No technological intervention identified |
| $^{15}O$ | Well-managed | None | None | No technological intervention identified |
| $^{103}Pd$ | Well-managed | None | None | Produced routinely using low-energy proton cyclotrons. |
| $^{223}Ra$ | At-risk | Investigate | Sustain | Expand $^{227}Ac$ generator feedstock using accelerator-driven thermal neutrons via $^{226}Ra(n,\gamma)^{227}Ra \rightarrow ^{227}Ac$. Investigate production using proton spallation via $^{nat}Th(p,x)^{227}Ac/^{227}Th/^{223}Ra$. |
| $^{82}Rb$ | At-risk | None | None | No technological intervention identified. |
| $^{75}Se$ | Well-managed | None | None | No technological intervention identified. |
| $^{89}Sr$ | Well-managed | None | None | No technological intervention identified. |
| $^{99m}Tc$ | At-risk | Established | Sustain | Supplement generator supply with regional proton-cyclotron production via $^{100}Mo(p,2n)^{99m}Tc$. *[More established]* |
| | | | | Supplement $^{99}$Mo/$^{99m}$Tc generator feedstock using accelerator-driven thermal neutrons via $^{235}$U(n,f)$^{99}$Mo. *[More investigative]* |
| $^{201}$Tl | At-risk | None | None | No technological intervention identified. |
| $^{133}$Xe | Well-managed | Investigate | Sustain | Produced routinely in reactors. Accelerator-driven thermal neutrons could provide a parallel supply route through $^{235}$U(n,f), followed by recovery of fission-product $^{133}$Xe. |
| $^{90}$Y | Well-managed | None | None | No technological intervention identified. |
| $^{89}$Zr | Well-managed | None | None | Produced using commercially available biomedical cyclotrons. |
| ***Underleveraged Middle*** | | | | |
| $^{198}$Au | Sig. constrained | Investigate | Expand | Carrier-added product is produced routinely in reactors. Investigate no-carrier-added $^{198}$Au production using $^{198}$Pt(p,n)$^{198}$Au, $^{198}$Pt(d,2n)$^{198}$Au, or $^{199}$Hg($\gamma$,p)$^{198}$Au, together with target recovery and chemical separation. |
| $^{169}$Er | Sig. constrained | Investigate | Expand | Carrier-added product is produced routinely in reactors. Investigate electromagnetic mass separation of $^{169}$Er produced through $^{168}$Er(n,$\gamma$)$^{169}$Er and accelerator-driven fast-neutron production via $^{169}$Tm(n,p)$^{169}$Er. |
| $^{186}$Re | Sig. constrained | Implement | Expand | Carrier-added product is produced routinely in reactors. No-carrier-added accelerator-based production via $^{186}$W(p,n)$^{186}$Re and $^{186}$W(d,2n)$^{186}$Re merits implementation and scale-up. |
| $^{188}$Re | Sig. constrained | Investigate | Expand | Established production requires rare, very high flux reactors. Investigate accelerator-based production via $^{nat}$Os($\gamma$, x)$^{188}$Re. |
| $^{153}$Sm | Sig. constrained | Investigate | Expand | Carrier-added product is produced routinely in reactors. No-carrier-added accelerator-based production via $^{150}$Nd($\alpha$,n)$^{153}$Sm merits investigation. |
| ***Emerging Frontier*** | | | | |
| $^{225}$Ac | Sig. constrained | Investigate | Enable | Investigate $^{226}$Ra ($\gamma$,n)$^{225}$Ra→$^{225}$Ac from electron linacs.<br>Investigate $^{226}$Ra (n,2n)$^{225}$Ra from accelerator fast neutrons. |
| $^{211}$At | Sig. constrained | Investigate | Enable | Primarily produced with alpha particle cyclotrons. Higher-current $\alpha$ beams and improved target cooling could scale established production via $^{209}$Bi($\alpha$,2n)$^{211}$At.<br>Heavy-ion production of $^{211}$Rn/$^{211}$At generators via $^{209}$Bi($^{7}$Li,5n)$^{211}$Rn→$^{211}$At merits investigation. |
| $^{213}$Bi | Sig. constrained | Investigate | Enable | Produced via $^{225}$Ac decay. See $^{225}$Ac assessment. |
| $^{67}$Cu | Sig. constrained | Established | Enable | High-purity photonuclear production at scale via $^{68}$Zn($\gamma$,p)$^{67}$Cu has been used for clinical trials. *[More established].* |
| | | | | Deuteron production via $^{70}$Zn(d,x)$^{67}$Cu and high-energy proton production via $^{68}$Zn(p,2p)$^{67}$Cu merit further development. *[More investigative]* |
| $^{212}$Pb | Sig. constrained | Investigate | Enable | Production using accelerator-driven fast neutrons via $^{226}$Ra(n,3n)$^{224}$Ra→$^{212}$Pb merits further investigation. Expansion of $^{228}$Th parent inventories with high-energy proton beams also merits investigation. |
| $^{44}$Sc | Sig. constrained | Established | Enable | Direct production using a low-energy proton cyclotron via $^{44}$Ca(p,n)$^{44}$Sc has been used for a clinical trial. *[More established]* |
| | | | | Higher-current proton accelerators could expand $^{44}$Ti/$^{44}$Sc generator feedstock via $^{45}$Sc(p,2n)$^{44}$Ti→$^{44}$Sc. Generator chemistry requires further investigation. *[More investigative]* |
| $^{47}$Sc | Sig. constrained | Investigate | Enable | Compare parent production via $^{46}$Ca(n,γ)$^{47}$Ca→$^{47}$Sc and $^{48}$Ca(γ,n)$^{47}$Ca→$^{47}$Sc with direct proton production via $^{nat}$V(p,x)$^{47}$Sc and photonuclear production via $^{nat}$V(γ, x)$^{47}$Sc, all merit investigation. |
| $^{149}$Tb | Sig. constrained | Investigate | Enable | Investigate production using proton spallation, via $^{165}$Ho(p,x)$^{149}$Tb or $^{nat}$Ta(p,x)$^{149}$Tb , followed by on-line mass separation. Investigate production using $^{12}$C ion beams via $^{142}$Nd($^{12}$C,$^{5n}$)$^{149}$Dy→$^{149}$Tb. Routes using $^{152}$Gd are bottlenecked by Gd enrichment. |
| $^{152}$Tb | Sig. constrained | Investigate | Enable | Investigate production using proton spallation, via $^{nat}$Ta(p,x)$^{152}$Tb. Investigate production using high energy protons, via $^{155}$Gd(p,$^{4n}$)$^{152}$Tb. Investigate production using $^{12}$C ions, via $^{nat}$Nd($^{12}$C,xn)$^{152}$Dy→$^{152}$Tb. |
| $^{161}$Tb | Sig. constrained | None | None | Current supply depends on enriched $^{160}$Gd irradiation and specialised Gd/Tb separation; no distinct unused technological intervention was identified. |
| $^{227}$Th | Sig. constrained | Investigate | Enable | Investigate production using proton spallation via $^{nat}$Th(p,x)$^{227}$Ac/$^{227}$Th. |

Table C1: A synthetic overview of the constraints, opportunities, and possible benefits associated with each isotope.

# References

1 National Nuclear Data Center. NuDat 3.0 database. https://www.nndc.bnl.gov/nudat3/.

2 International Atomic Energy Agency. Central nervous system radiotracer development: bench to bedside. Vienna: International Atomic Energy Agency, 2024 https://www.iaea.org/publications/15682/central-nervous-system-radiotracer-development-bench-to-bedside.

3 Galldiks N, Lohmann P, Aboian M, *et al*. Update to the RANO working group and EANO recommendations for the clinical use of PET imaging in gliomas. *Lancet Oncol* 2025; **26**: e436–47.

4 Mayo C. Choline C 11 injection: prescribing information. Rochester (MN): Mayo Clinic, 2012 https://dailymed.nlm.nih.gov/dailymed/drugInfo.cfm?setid=61f00cb4-ea53-4ba6-a890-78586095d2f6.

5 International Committee for Standardization in Haematology. Recommended methods for measurement of red-cell and plasma volume. *J Nucl Med* 1980; **21**: 793–800.

6 Manual for reactor produced radioisotopes. International Atomic Energy Agency, 2003.

7 International Atomic Energy Agency. Selecting megavoltage treatment technologies in external beam radiotherapy. 2022.

8 International Atomic Energy Agency. Trends in radiation sterilization of health care products. 2008.

9 Ahmedova A, Todorov B, Burdzhiev N, Goze C. Copper radiopharmaceuticals for theranostic applications. *European Journal of Medicinal Chemistry* 2018; **157**: 1406–25.

10 Mikolajczak R, Van Der Meulen NP, Lapi SE. Radiometals for imaging and theranostics, current production, and future perspectives. *Labelled Comp Radiopharmac* 2019; **62**: 615–34.

11 Adam Rebeles R, Van Den Winkel P, Hermanne A, Tárkányi F. New measurement and evaluation of the excitation function of 64Ni(p,n) reaction for the production of 64Cu. *Nuclear Instruments and Methods in Physics Research Section B: Beam Interactions with Materials and Atoms* 2009; **267**: 457–61.

12 International Atomic Energy Agency. Production and quality control of 18F labelled radiopharmaceuticals. 2021.

13 Boellaard R, Delgado-Bolton R, Oyen W, *et al*. FDG PET/CT: EANM procedure guidelines for tumour imaging: version 2.0. *Eur J Nucl Med Mol Imaging* 2015; **42**: 328–54.

14 US Food and Drug Administration. PET diagnostic radiopharmaceutical drugs: Medical Imaging Drugs Advisory Committee briefing document. 2023.

15 GE Healthcare Inc. Flyrcado (flurpiridaz F 18) injection: prescribing information. Arlington Heights (IL): GE Healthcare Inc, 2025

https://dailymed.nlm.nih.gov/dailymed/lookup.cfm?setid=3bb6caf8-ac98-46c7-a9f0-961e66032705.

16 International Atomic Energy Agency. Cyclotron produced radioisotopes: physical characteristics and production methods. 2009.

17 Palestro C, Clark A, Grady E, *et al.* Appropriate use criteria for the use of nuclear medicine in musculoskeletal infection imaging. *J Nucl Med* 2021; **62**: 1815–31.

18 Abikhzer G, Treglia G, Pelletier-Galarneau M, others. EANM/SNMMI guideline/procedure standard for 18F-FDG hybrid PET use in infection and inflammation in adults, version 2.0. *Eur J Nucl Med Mol Imaging* 2025; **52**: 510–38.

19 US Food and Drug Administration. Locametz prescribing information. 2022.

20 US Food and Drug Administration. Netspot prescribing information. 2023.

21 Hope T, Calais J, Zhang L, Dieckmann W, Millo C. SNMMI procedure standard/EANM practice guideline for SSTR PET: imaging neuroendocrine tumors. *J Nucl Med* 2023; **64**: 204–10.

22 International Atomic Energy Agency. Good management practices of cyclotron based 68Ga radiopharmaceuticals. 2026.

23 Palestro C, Brown M, Forstrom L, others. Society of Nuclear Medicine procedure guideline for 111In-leukocyte scintigraphy for suspected infection/inflammation, version 3.0. 2004. https://snmmi.org/Web/Web/Clinical-Practice/Procedure-Standards/Standards/111In-Leukocyte-Scintigraphy-for-Suspected-Infection-Inflammation-3-0.aspx.

24 Kumar K, Ghosh A. Radiochemistry, Production Processes, Labeling Methods, and ImmunoPET Imaging Pharmaceuticals of Iodine-124. *Molecules* 2021; **26**: 414.

25 Hermanne A, Tarkanyi F, Takacs S, *et al.* Limitation of the long-lived 121Te contaminant in production of 123I through the 124Xe(p,x) route. *Applied Radiation and Isotopes* 2011; **69**: 358–68.

26 Henry A, Pieters B, Siebert F, Hoskin P. GEC-ESTRO ACROP prostate brachytherapy guidelines. *Radiother Oncol* 2022; **167**: 244–51.

27 International Atomic Energy Agency. Production techniques and quality control of sealed radioactive sources of 103Pd, 125I, 192Ir and ytterbium-169. 2006.

28 Collaborative Ocular Melanoma Study Group. The COMS randomized trial of 125I brachytherapy for choroidal melanoma, III. *Arch Ophthalmol* 2001; **119**: 969–82.

29 Qaim S. Cyclotron production of medical radionuclides. In: Handbook of nuclear chemistry. Boston (MA): Springer, 2011: 1903–33.

30 International Atomic Energy Agency. Optimization in brachytherapy. https://www.iaea.org/resources/rpop/health-professionals/radiotherapy/brachytherapy/optimization.

31 Qaim S. Medical radioisotope production: science and technology. 2020. DOI:10.1515/9783110604375.

32 Bajc M, Schümichen C, Grüning T, others. EANM guideline for ventilation/perfusion SPECT for diagnosis of pulmonary embolism and beyond. *Eur J Nucl Med Mol Imaging* 2019; **46**: 2429–51.

33 Medicine S of N, Molecular I, European Association of Nuclear M, American College of Nuclear M, Australian, New Zealand Society of Nuclear M. SNMMI/EANM/ACNM/ANZSNM guideline for ventilation-perfusion pulmonary scintigraphy. 2026. https://snmmi.org/Web/Web/Clinical-Practice/Procedure-Standards/Procedure-Standards.aspx.

34 Banerjee S, Pillai MRA, Knapp FF (Russ). Lutetium-177 Therapeutic Radiopharmaceuticals: Linking Chemistry, Radiochemistry, and Practical Applications. *Chem Rev* 2015; **115**: 2934–74.

35 Advanced Accelerator Applications USA Inc. Lutathera (lutetium Lu 177 dotatate) injection: prescribing information. Millburn (NJ): Advanced Accelerator Applications USA Inc, 2024 https://dailymed.nlm.nih.gov/dailymed/drugInfo.cfm?setid=72d1a024-00b7-418a-b36e-b2cb48f2ab55.

36 Nagai Y, Kawabata M, Hashimoto S, *et al.* Estimated Isotopic Compositions of Yb in Enriched $^{176}$Yb for Producing $^{177}$Lu with High Radionuclide Purity by $^{176}$Yb($d$, $x$)$^{177}$Lu. *J Phys Soc Jpn* 2022; **91**: 044201.

37 US Food and Drug Administration. Ammonia N 13 injection prescribing information. 2011.

38 Packard R, Maddahi J, Pelletier-Galarneau M, *et al.* SNMMI/EANM/ASNC/ACNM procedure standard/practice guideline for 18F-flurpiridaz PET myocardial perfusion imaging and blood flow quantitation. *J Nucl Med* 2025; **66**: 1538–54.

39 Danad I, Szymonifka J, Twisk J, others. Comparison of coronary CT angiography, SPECT, PET, and hybrid imaging for diagnosis of ischemic heart disease determined by fractional flow reserve. *JAMA Cardiol* 2017; **2**: 1100–7.

40 Powell J, O'Neil J. Production of [15O]water at low-energy proton cyclotrons. *Appl Radiat Isot* 2006; **64**: 755–9.

41 Queern SL, Cardman R, Loveless CS, Shepherd MR, Lapi SE. Production of $^{15}$O for Medical Applications via the $^{16}$O(γ,n)$^{15}$O Reaction. *J Nucl Med* 2019; **60**: 424–8.

42 Skowronek J. Current status of brachytherapy in cancer treatment – short overview. *jcb* 2017; **9**: 581–9.

43 Bayer HealthCare Pharmaceuticals Inc. Xofigo (radium Ra 223 dichloride) injection: prescribing information. 2026. https://dailymed.nlm.nih.gov/dailymed/lookup.cfm?setid=a398400e-bd31-41a9-9696-4f7c06569ede.

44 Parker C, Nilsson S, Heinrich D, others. Alpha emitter radium-223 and survival in metastatic prostate cancer. *N Engl J Med* 2013; **369**: 213–23.

45 Radchenko V, Morgenstern A, Jalilian A, *et al.* Production and supply of alpha-particle-emitting radioisotopes for targeted alpha-therapy. *J Nucl Med* 2021; **62**: 1495–503.

46 American Society of Nuclear Cardiology. PET myocardial blood flow practice points. 2025.

47 US Food and Drug Administration. CardioGen-82 prescribing information. 2026.

48 Murthy V, Bateman T, Beanlands R, others. Clinical quantification of myocardial blood flow using PET. *J Nucl Med* 2018; **59**: 273–93.

49 Westwood M, Corro Ramos I, Armstrong N, *et al.* SeHCAT (tauroselcholic [75Se] acid) for the investigation of bile acid diarrhoea in adults: a systematic review and cost-effectiveness analysis. *Health Technol Assess* 2022; **26**. DOI:10.3310/JTFO0945.

50 Medicines Evaluation Board. SeHCAT 370 kBq capsules: public assessment report. 2024.

51 Amjed N, Aslam M, Hussain M, Qaim S. Evaluation of nuclear reaction cross section data of proton- and deuteron-induced reactions on 75As, with particular emphasis on the production of 73Se. *Radiochim Acta* 2021; **109**: 525–37.

52 Lewington V, McEwan A, Ackery D, *et al.* A prospective, randomised double-blind crossover study to examine the efficacy of strontium-89 in pain palliation in patients with advanced prostate cancer metastatic to bone. *Eur J Cancer* 1991; **27**: 954–8.

53 US Department of Energy Isotope P, National Isotope Development C. Ensuring a steady supply of 89Sr. 2024. https://www.isotopes.gov/Ensuring-steady-supply-of-strontium-89.

54 Spahn I, Coenen H, Qaim S. Enhanced production possibility of the therapeutic radioisotopes 64Cu, 67Cu and 89Sr via (n,p) reactions induced by fast spectral neutrons. *Radiochim Acta* 2004; **92**: 183–6.

55 Nuclear Energy Agency. Current trends in the supply and utilisation of medical radioisotopes. 2025. https://doi.org/10.82155/x9kb-7j83.

56 Chiesa C, Sjögren-Gleisner K, Walrand S, *et al.* EANM dosimetry committee series on standard operational procedures: a unified methodology for 99mTc-MAA pre- and 90Y peri-therapy dosimetry in liver radioembolization with 90Y microspheres. *EJNMMI Phys* 2021; **8**: 77.

57 Ma M. Receptor-targeted 99mTc radiopharmaceuticals: increasing access to molecular imaging in healthcare. *Front Chem Biol* 2026; **4**: 1729615.

58 International Atomic Energy Agency. Alternative radioisotope production with a cyclotron. 2021. https://www-pub.iaea.org/MTCD/Publications/PDF/P1937_web.pdf.

59 Curium US LLC. Thallous chloride Tl 201 injection: prescribing information. 2025. https://dailymed.nlm.nih.gov/dailymed/drugInfo.cfm?setid=0da81fc0-a137-46d3-9378-d8053a82e61a.

60 Curium US LLC. Xenon Xe 133 gas: prescribing information. 2024. https://dailymed.nlm.nih.gov/dailymed/drugInfo.cfm?setid=5eb971be-8808-4aeb-9898-0d22d5dffe04.

61 Acrotech Biopharma Inc. Zevalin (ibritumomab tiuxetan) injection: prescribing information. 2023.

62 Sirtex Medical Inc. SIR-Spheres 90Y resin microspheres: instructions for use. 2025.

63 Chakravarty R, Dash A, Pillai M. Availability of 90Y from 90Sr: a nuclear medicine perspective. *Cancer Biother Radiopharm* 2012; **27**: 621–41.

64 International Atomic Energy Agency. Therapeutic radionuclide generators: 90Sr/90Y and 188W/188Re generators. Vienna: International Atomic Energy Agency, 2009 https://www.iaea.org/publications/8045/therapeutic-radionuclide-generators-90sr90y-and-188w188re-generators.

65 Korde A. New coordinated research project: 89Zr production and 89Zr radiopharmaceuticals (F22071). 2018.

66 Jauw Y, Menke-van der Houven van Oordt C, Hoekstra O, *et al.* Immuno-positron emission tomography with 89Zr-labeled monoclonal antibodies in oncology: what can we learn from initial clinical trials? Front Pharmacol. 2016. DOI:10.3389/fphar.2016.00131.

67 Wooten A, Madrid E, Schweitzer G, *et al.* Routine production of 89Zr using an automated module. *Appl Sci* 2013; **3**: 593–613.

68 Kampen W, Boddenberg-Pätzold B, Fischer M, *et al.* The EANM guideline for radiosynoviorthesis. *Eur J Nucl Med Mol Imaging* 2022; **49**: 681–708.

69 Sadler AWE, Hogan L, Fraser B, Rendina LM. Cutting edge rare earth radiometals: prospects for cancer theranostics. *EJNMMI radiopharm chem* 2022; **7**: 21.

70 Talip Z, Borgna F, Müller C, *et al.* Production of Mass-Separated Erbium-169 Towards the First Preclinical in vitro Investigations. *Front Med* 2021; **8**: 643175.

71 Bhardwaj P, Claude G, Brenner W, Spreckelmeyer S. Radioactive Gold Isotopes in the Era of Theranostics: Medical Potential of $^{198}$Au and $^{199}$Au Radioisotopes. *Nuklearmedizin* 2026; **65**: 69–82.

72 Quirijnen J, Han S, Zonnenberg B, others. Efficacy of 186Re-etidronate in prostate cancer patients with metastatic bone pain. *J Nucl Med* 1996; **37**: 1511–5.

73 Uccelli L, Martini P, Pasquali M, Boschi A. Rhenium radioisotopes for medicine: production and applications. *Molecules* 2022; **27**: 5283.

74 Mastren T, Radchenko V, Hopkins P, others. Bulk production and evaluation of high-specific-activity 186Re using enriched 186W targets. *Nucl Med Biol* 2017; **49**: 24–9.

75 Shinto A, Mallia M, Kameswaran M, others. Clinical utility of 188Re-hydroxyethylidene-1,1-diphosphonate as a bone pain palliative in multiple malignancies. *World J Nucl Med* 2018; **17**: 228–35.

76 Bernal P, Raoul J, Vidmar G, *et al.* Intra-arterial rhenium-188 lipiodol in the treatment of inoperable hepatocellular carcinoma: results of an IAEA-sponsored multinational study. *Int J Radiat Oncol Biol Phys* 2007; **69**: 1448–55.

77 Cardaci G, Baxi S, Vohra S, *et al.* Efficacy, safety, and patient reported outcomes of rhenium-skin cancer therapy for non-melanoma skin cancer: 1-year results from the EPIC-Skin study. *Adv Radiat Oncol* 2025; **10**: 101802.

78 Ersöz O, Spink R, Griswold J, Yurt F, Mirzadeh S. Measurement of neutron capture cross section of 187W for production of 188W. *Appl Radiat Isot* 2019; **148**: 191–6.

79 Lantheus Medical Imaging Inc. Quadramet (samarium Sm 153 lexidronam) injection: prescribing information. 2017.

80 Vermeulen K, Van de Voorde M, Segers C, *et al.* Exploring the potential of high-molar-activity samarium-153 for targeted radionuclide therapy with [153Sm]Sm-DOTA-TATE. *Pharmaceutics* 2022; **14**: 2566.

81 Aikawa M, Sakaguchi M, Ukon N, *et al.* Production cross sections of samarium-153 and -145 via alpha-particle-induced reactions on natural neodymium. *Appl Radiat Isot* 2022; **187**: 110345.

82 Kratochwil C, Bruchertseifer F, Rathke H, *et al.* Targeted alpha-therapy of metastatic castration-resistant prostate cancer with 225Ac-PSMA-617: dosimetry estimate and empiric dose finding. *J Nucl Med* 2017; **58**: 1624–31.

83 International Atomic Energy Agency. Production and quality control of 225Ac radiopharmaceuticals. 2024.

84 Diehl G, Holschuh TV, Foley-Janish A, *et al.* Photonuclear production of 225Ac from 226Ra targets. *J Radioanal Nucl Chem* 2025; **334**: 6209–17.

85 Morrell J. Next-generation isotope production via deuteron breakup. 2021. https://escholarship.org/uc/item/1cj6716s.

86 Morrell J, Voyles A, Batchelder J, Brown J, Bernstein L. Secondary neutron production from thick-target deuteron breakup. *Phys Rev C* 2023; **108**: 024616.

87 Albertsson P, Bäck T, Bergmark K, *et al.* Astatine-211 based radionuclide therapy: Current clinical trial landscape. *Front Med* 2023; **9**: 1076210.

88 Müller M, Pedersen NB, Shalgunov V, Jensen AI, Battisti UM, Herth MM. Astatine-211—Towards In Vivo Stable Astatine-211 Labeled Radiopharmaceuticals and Their (Pre)Clinical Applications. *Medicinal Research Reviews* 2026; **46**: 203–37.

89 Watabe T, Hatano K, Naka S, others. First-in-human SPECT/CT imaging of [211At]PSMA-5: targeted alpha therapy in a patient with refractory prostate cancer. *Eur J Nucl Med Mol Imaging* 2025; **52**: 2253–5.

90 Feng Y, Zalutsky MR. Production, purification and availability of 211At: Near term steps towards global access. *Nuclear Medicine and Biology* 2021; **100–101**: 12–23.

91 Zimmermann R. Are$^{61}$ Cu,$^{64}$ Cu, and$^{67}$ Cu Really Happening? *J Nucl Med* 2026; **67**: 1032–5.

92 Nigron E, Guertin A, Haddad F, Sounalet T. Is 70Zn(d,x)67Cu the Best Way to Produce 67Cu for Medical Applications? *Front Med* 2021; **8**: 674617.

93 Delpassand E, Tworowska I, Esfandiari R, *et al.* Targeted alpha-emitter therapy with 212Pb-DOTAMTATE for the treatment of metastatic SSTR-expressing neuroendocrine tumors: first-in-humans dose-escalation clinical trial. *J Nucl Med* 2022; **63**: 1326–33.

94 Meredith R, Torgue J, Rozgaja T, *et al.* Safety and outcome measures of first-in-human intraperitoneal alpha radioimmunotherapy with 212Pb-TCMC-trastuzumab. *Am J Clin Oncol* 2018; **41**: 716–21.

95 Berner K, Hernes E, Kvassheim M, *et al.* First-in-human phase 0 study of AB001, a prostate-specific membrane antigen-targeted 212Pb radioligand, in patients with metastatic castration-resistant prostate cancer. *J Nucl Med* 2025; **66**: 732–8.

96 Scaffidi-Muta J, Abell A. 212Pb in targeted radionuclide therapy: a review. *EJNMMI Radiopharm Chem* 2025; **10**: 34.

97 McNeil B, Robertson A, Fu W, *et al.* Production, purification, and radiolabeling of the 203Pb/212Pb theranostic pair. *EJNMMI Radiopharm Chem* 2021; **6**: 6.

98 Singh A, Van Der Meulen NP, Müller C, *et al.* First-in-Human PET/CT Imaging of Metastatic Neuroendocrine Neoplasms with Cyclotron-Produced$^{44}$ Sc-DOTATOC: A Proof-of-Concept Study. *Cancer Biotherapy and Radiopharmaceuticals* 2017; **32**: 124–32.

99 Tashima H, Yamaya T. Three-Gamma Imaging in Nuclear Medicine: A Review. *IEEE Trans Radiat Plasma Med Sci* 2024; **8**: 853–66.

100 Cingoranelli S, Appiah J, Putnam E, Lapi S. Production of high-purity 47Sc from proton irradiation of natural vanadium targets. *EJNMMI Radiopharm Chem* 2024; **9**: 89.

101 Beyer G, Miederer M, Vranješ-Đurić S, *et al.* Targeted alpha therapy in vivo: direct evidence for single cancer cell kill using 149Tb-rituximab. *Eur J Nucl Med Mol Imaging* 2004; **31**: 547–54.

102 Müller C, Singh A, Umbricht C, *et al.* Preclinical investigations and first-in-human application of 152Tb-PSMA-617 for PET/CT imaging of prostate cancer. *EJNMMI Res* 2019; **9**: 68.

103 Umbricht C, Köster U, Bernhardt P, *et al.* Alpha-PET for prostate cancer: preclinical investigation using 149Tb-PSMA-617. *Sci Rep* 2019; **9**: 17800.

104 Favaretto C, Grundler P, Talip Z, *et al.* 149Tb production: a focus on yield and quality improvement towards preclinical application. *Sci Rep* 2024; **14**: 3284.

105 Baum R, Singh A, Kulkarni H, *et al.* First-in-humans application of 161Tb: a feasibility study using 161Tb-DOTATOC. *J Nucl Med* 2021; **62**: 1391–7.

106 Allen B, Goozee G, Sarkar S, Beyer G, Morel C, Byrne A. Production of terbium-152 by heavy-ion reactions and proton-induced spallation. *Appl Radiat Isot* 2001; **54**: 53–8.

107 Köster U, Assmann W, Bacri C, *et al.* Electromagnetic isotope separation of gadolinium isotopes for the production of 152,155Tb for radiopharmaceutical applications. *Nucl Instrum Methods Phys Res B* 2020; **463**: 111–4.

108 Buteau J, Kostos L, Jackson P, *et al.* First-in-human results of 161Tb-PSMA-I&T dual beta-Auger radioligand therapy in patients with metastatic castration-resistant prostate cancer (VIOLET): a single-centre, single-arm, phase 1/2 study. *Lancet Oncol* 2025; **26**: 1009–17.

109 Gracheva N, Müller C, Talip Z, *et al.* Production and characterization of no-carrier-added 161Tb as an alternative to the clinically applied 177Lu for radioisotope therapy. *EJNMMI Radiopharm Chem* 2019; **4**: 12.

110 Karlsson J, Schatz C, Wengner A, *et al.* Targeted 227Th conjugates as treatment options in oncology. *Front Med (Lausanne)* 2023; **9**: 1071086.

111 Minchom A, Lindén O, Knapen D, *et al.* First-in-human study of a 227Th-labeled mesothelin-targeting antibody-chelator conjugate in patients with malignant mesothelioma and other solid tumors. *ESMO Open* 2025; **10**: 105905.

112 Griswold J, Medvedev D, Engle J, *et al.* Large-scale accelerator production of 225Ac: effective cross sections for 78-192 MeV protons incident on 232Th targets. *Appl Radiat Isot* 2016; **118**: 366–74.

113 Qaim SM. Nuclear data for production and medical application of radionuclides: Present status and future needs. *Nuclear Medicine and Biology* 2017; **44**: 31–49.

114 Tárkányi FT, Ignatyuk AV, Hermanne A, *et al.* Recommended nuclear data for medical radioisotope production: diagnostic gamma emitters. *J Radioanal Nucl Chem* 2019; **319**: 487–531.

115 Cingoranelli SJ, Burnett L, Putnam E, Lapi SE. Cross section measurements for the production of 49,51Cr and 47Sc from proton irradiation of natural vanadium up to 24 MeV. *Applied Radiation and Isotopes* 2024; **212**: 111469.

116 Tárkányi F, Ditrói F, Takács S, Hermanne A, Baba M, Ignatyuk AV. Investigation of activation cross-sections of deuteron induced reactions on vanadium up to 40 MeV. *Nuclear Instruments* *and Methods in Physics Research Section B: Beam Interactions with Materials and Atoms* 2011; **269**: 1792–800.

117 Aslam M, Sudár S, Hussain M, Malik AA, Shah HA, Qaim SM. Charged particle induced reaction cross section data for production of the emerging medically important positron emitter$^{64}$ Cu: A comprehensive evaluation. *Radiochimica Acta* 2009; **97**: 669–86.

118 Kozempel J, Abbas K, Simonelli F, *et al.* A novel method for n.c.a.$^{64}$ Cu production by the$^{64}$ Zn( *d* , 2 *p* )$^{64}$ Cu reaction and dual ion-exchange column chromatography. *Radiochimica Acta* 2007; **95**: 75–80.

119 Al-Azraq BT, Mahmood RI, Mahmood LA, Mohsin RCh, Jaffer RS. Influence of level density models on proton and deuteron-induced reactions using zinc target for the production of 66-68Ga medical radioisotopes. *Annals of Nuclear Energy* 2025; **220**: 111535.

120 IAEA. Alternative Radionuclide Production with a Cyclotron. Vienna: IAEA, 2021.

121 Internationale Atomenergie-Organisation, editor. Production of long lived parent radionuclides for generators: 68Ge, 82Sr, 90Sr and 188W. Vienna: IAEA, 2010.

122 Tárkányi FT, Ignatyuk AV, Hermanne A, *et al.* Recommended nuclear data for medical radioisotope production: diagnostic positron emitters. *J Radioanal Nucl Chem* 2019; **319**: 533–666.

123 IAEA. Gallium-68 Cyclotron Production. Vienna: IAEA, 2019.

124 Nikjou A, Sadeghi M, Sharifian M, Baghbani R. Nuclear model calculations on the production of auger electron emitter 111In: As a theranostic radionuclide. *Applied Radiation and Isotopes* 2020; **166**: 109354.

125 Hermanne A, Tarkanyi F, Takacs S, *et al.* Limitation of the long-lived 121Te contaminant in production of 123I through the 124Xe(p,x) route. *Applied Radiation and Isotopes* 2011; **69**: 358–68.

126 Avetisyan A, Avagyan R, Dallakyan R, *et al.* Investigation of 123 I production using electron accelerator. *Nuclear Medicine and Biology* 2017; **47**: 44–7.

127 Bastian T, Coenen HH, Qaim SM. Excitation functions of 124Te(d,xn)124,125I reactions from threshold up to 14MeV: comparative evaluation of nuclear routes for the production of 124I. *Applied Radiation and Isotopes* 2001; **55**: 303–8.

128 Aslam MN, Sudár S, Hussain M, Malik AA, Qaim SM. Evaluation of excitation functions of 3He- and α-particle induced reactions on antimony isotopes with special relevance to the production of iodine-124. *Applied Radiation and Isotopes* 2011; **69**: 94–104.

129 Qaim SM, editor. Nuclear data for the production of therapeutic radionuclides. Vienna: Internat. Atomic Energy Agency, 2011.

130 Hilgers K, Sudár S, Qaim SM. Experimental study and nuclear model calculations on the 192Os(p,n)192Ir reaction: Comparison of reactor and cyclotron production of the therapeutic radionuclide 192Ir. *Applied Radiation and Isotopes* 2005; **63**: 93–8.

131 Tárkányi F, Hermanne A, Takács S, *et al*. Study of the 192Os(d,2n) reaction for production of the therapeutic radionuclide 192Ir in no-carrier added form. *Applied Radiation and Isotopes* 2007; **65**: 1215–20.

132 Indira G, Anbalagan G. Investigation of the production routes of $^{81,82m,83,84}$ Rb isotopes through proton-induced nuclear reactions on $^{82,83,84,86}$ Kr. *J Phys G: Nucl Part Phys* 2024; **51**: 115101.

133 Alrumayan F, Okarvi SM, Nagatsu K, Yanbawi S, Aljammaz I. Development of a nickel plated aluminum krypton-81m target system. *Applied Radiation and Isotopes* 2017; **121**: 6–11.

134 Dash A, Pillai MRA, Knapp FF. Production of 177Lu for Targeted Radionuclide Therapy: Available Options. *Nucl Med Mol Imaging* 2015; **49**: 85–107.

135 Nagai Y, Kawabata M, Hashimoto S, *et al.* Estimated isotopic compositions of Yb in enriched 176Yb for producing 177Lu with high radionuclide purity by 176Yb(d,x)177Lu. *J Phys Soc Jpn* 2022; **91**: 044201.

136 Shao L. Optimization of Deuteron Irradiation of 176Yb for Producing 177Lu of High Specific Activity Exceeding 3000 GBq/mg. *Molecules* 2023; **28**: 6053.

137 Kazakov AG, Belyshev SS, Ekatova TYu, Khankin VV, Kuznetsov AA, Aliev RA. Production of 177Lu by hafnium irradiation using 55-MeV bremsstrahlung photons. *J Radioanal Nucl Chem* 2018; **317**: 1469–76.

138 Internationale Atomenergie-Organisation, editor. Cyclotron produced radionuclides: operation and maintenance of gas and liquid targets. Vienna: Internat. Atomic Energy Agency, 2012.

139 Szucs Z, Hamkens W, Takács S, Tárkányi F, Coenen HH, Qaim SM. Excitation Functions of 14N(d,t)13N and 14N(d,«n)uC Reactions from Threshold to 12.3 MeV : Radionuclidic Purity of 150 Produced via the 14N(d,n)lsO Reaction. .

140 Mulholland GK, Kilbourn MR, Moskwa JJ. Direct Simultaneous Production of [ 50]Water and [ 3N]Ammonia or [ 8F]Fluoride Ion by 26 MeV Proton Irradiation of a Double Chamber Water Target. .

141 Hermanne A, Sonck M, Fenyvesi A, Daraban L. Study on production of 103Pd and characterisation of possible contaminants in the proton irradiation of 103Rh up to 28 Mev. *Nuclear Instruments and Methods in Physics Research Section B: Beam Interactions with Materials and Atoms* 2000; **170**: 281–92.

142 Manenti S, Alí Santoro MDC, Cotogno G, *et al.* Excitation function and yield for the 103Rh(d,2n)103Pd nuclear reaction: Optimization of the production of palladium-103. *Nuclear Medicine and Biology* 2017; **49**: 30–7.

143 Hussain M, Sudar S, Aslam MN, *et al.* A comprehensive evaluation of charged-particle data for production of the therapeutic radionuclide 103Pd. *Applied Radiation and Isotopes* 2009; **67**: 1842–54.

144 Dash A, Das T. Production of therapeutic radionuclides for bone-seeking radiopharmaceuticals to meet clinical exigency and academic research. *J Radioanal Nucl Chem* 2025; **334**: 2591–611.

145 Weidner JW, Mashnik SG, John KD, *et al.* 225Ac and 223Ra production via 800MeV proton irradiation of natural thorium targets. *Applied Radiation and Isotopes* 2012; **70**: 2590–5.

146 Z. Soderquist C, K. McNamara B, R. Fisher D. Production of High-purity Radium-223 from Legacy Actinium-Beryllium Neutron Sources. *CRP* 2012; **5**: 244–52.

147 Dearmon HD, Krane KS. Neutron capture cross sections of 74, 76, 78, 80, 82Se. *Eur Phys J A* 2019; **55**: 135.

148 Amjed N, Aslam MN, Hussain M, Qaim SM. Evaluation of nuclear reaction cross section data of proton and deuteron induced reactions on$^{75}$ As, with particular emphasis on the production of$^{73}$ Se. *Radiochimica Acta* 2021; **109**: 525–37.

149 Das T, Pillai MRA. Options to meet the future global demand of radionuclides for radionuclide therapy. *Nuclear Medicine and Biology* 2013; **40**: 23–32.

150 OECD, Agency NE. The Supply of Medical Isotopes: An Economic Diagnosis and Possible Solutions. OECD. 2019; published online Nov 17. DOI:10.1787/9b326195-en.

151 Tárkányi F, Hermanne A, Takács S, Shubin YuN, Dityuk AI. Cross sections for production of the therapeutic radioisotopes$^{198}$ Au and$^{199}$ Au in proton and deuteron induced reactions on$^{198}$ Pt. *Radiochimica Acta* 2004; **92**: 223–8.

152 Khandaker MU, Haba H, Kassim HA. Cyclotron produced 198gAu, a potential radionuclide for diagnostic and therapeutic applications. Bandung, Indonesia, 2016: 080001.

153 Hussain M, Sudár S, Aslam MN, Malik AA, Ahmad R, Qaim SM. Evaluation of charged particle induced reaction cross section data for production of the important therapeutic radionuclide$^{186}$ Re. *Radiochimica Acta* 2010; **98**: 385–95.

154 Zheltonozhskaya MV, Remizov PD, Lenivkin MV, Kuzmenkov AA, Khankin VV, Chernyaev AP. Study of the possibility of producing rhenium medical isotopes at electron accelerators. *Eur Phys J Spec Top* 2026; **235**: 1405–12.

155 Qaim SM, Spahn I, Kandil SA, Coenen HH. Nuclear data for production of$^{88}$ Y,$^{140}$ Nd,$^{153}$ Sm and$^{169}$ Yb *via* novel routes. *Radiochimica Acta* 2007; **95**: 313–7.

156 Sasaki Y, Sano A, Sasaki S, *et al.* Evaluation of the production amount of$^{225}$ Ac and its uncertainty through the$^{226}$ Ra(n,2n) reaction in the experimental fast reactor Joyo. *Journal of Nuclear Science and Technology* 2024; **61**: 509–20.

157 Bernstein LA, Voyles AS. Simultaneous Actinium-225 and Lead-212 Production using Secondary Neutrons. *Journal of Medical Imaging and Radiation Sciences* 2026; **57**: 102259.

158 Maeda E, Yokoyama A, Taniguchi T, Washiyama K, Nishinaka I. Measurements of the excitation functions of radon and astatine isotopes from 7Li-induced reactions with 209Bi for development of a 211Rn–211At generator. *J Radioanal Nucl Chem* 2020; **323**: 921–6.

159 Aliev RA, Belyshev SS, Kuznetsov AA, *et al.* Photonuclear production and radiochemical separation of medically relevant radionuclides: 67Cu. *J Radioanal Nucl Chem* 2019; **321**: 125–32.

160 Qaim S. New directions in nuclear data research for accelerator-based production of medical radioisotopes. *J Radioanal Nucl Chem* 2024; **333**: 3577–84.

161 Medvedev DG, Mausner LF, Meinken GE, *et al.* Development of a large scale production of 67Cu from 68Zn at the high energy proton accelerator: Closing the 68Zn cycle. *Applied Radiation and Isotopes* 2012; **70**: 423–9.

162 Schmidt CE, Gajecki L, Deri MA, Sanders VA. Current State of 44Ti/44Sc Radionuclide Generator Systems and SeparationChemistry. *CRP* 2023; **16**: 95–106.

163 Dellepiane G, Casolaro P, Gottstein A, Mateu I, Scampoli P, Braccini S. 44Sc production from enriched 47TiO2 targets with a medical cyclotron. *Applied Radiation and Isotopes* 2024; **206**: 111220.

164 Domnanich KA, Müller C, Benešová M, *et al.* 47Sc as useful β−-emitter for the radiotheragnostic paradigm: a comparative study of feasible production routes. *EJNMMI radiopharm chem* 2017; **2**: 5.

165 Starovoitova VN, Cole PL, Grimm TL. Accelerator-based photoproduction of promising beta-emitters 67Cu and 47Sc. *J Radioanal Nucl Chem* 2015; **305**: 127–32.

166 Pupillo G, Mou L, Boschi A, *et al.* Production of 47Sc with natural vanadium targets: results of the PASTA project. *J Radioanal Nucl Chem* 2019; **322**: 1711–8.

167 Nardo LD, Pupillo G, Mou L, *et al.* Preliminary dosimetric analysis of DOTA-folate radiopharmaceutical radiolabelled with$^{47}$ Sc produced through$^{nat}$ V(p,x)$^{47}$ Sc cyclotron irradiation. *Phys Med Biol* 2021; **66**: 025003.

168 Snow MS, Foley A, Ward JL, Kinlaw MT, Stoner J, Carney KP. High purity 47Sc production using high-energy photons and natural vanadium targets. *Applied Radiation and Isotopes* 2021; **178**: 109934.

169 Mamtimin M, Harmon F, Starovoitova VN. Sc-47 production from titanium targets using electron linacs. *Applied Radiation and Isotopes* 2015; **102**: 1–4.

170 Misiak R, Walczak R, Wąs B, Bartyzel M, Mietelski JW, Bilewicz A. 47Sc production development by cyclotron irradiation of 48Ca. *J Radioanal Nucl Chem* 2017; **313**: 429–34.

171 Dellepiane G, Casolaro P, Mateu I, Scampoli P, Voeten N, Braccini S. 47Sc and 46Sc cross-section measurement for an optimized 47Sc production with an 18 MeV medical PET cyclotron. *Applied Radiation and Isotopes* 2022; **189**: 110428.

172 Beyer GJ, Čomor JJ, Daković M, *et al.* Production routes of the alpha emitting$^{149}$ Tb for medical application. *Radiochimica Acta* 2002; **90**: 247–52.

173 Steyn GF, Vermeulen C, Szelecsényi F, *et al.* Cross sections of proton-induced reactions on 152Gd, 155Gd and 159Tb with emphasis on the production of selected Tb radionuclides. *Nuclear Instruments and Methods in Physics Research Section B: Beam Interactions with Materials and Atoms* 2014; **319**: 128–40.

174 Naskar N, Lahiri S. Theranostic Terbium Radioisotopes: Challenges in Production for Clinical Application. *Front Med* 2021; **8**: 675014.